\documentclass[preprint,times]{aastex63}
\newcommand \versnum {26}
\usepackage{color}
\usepackage{natbib}
\usepackage{amsmath}
\usepackage{nicefrac}
\tightenlines

\newcommand \bmu        {{\boldsymbol{\mu}}}

\newcommand \bahat      {\hat{\bf a}}
\newcommand \bB         {{\bf B}}

\newcommand \bBhat      {\hat{\bf B}}

\newcommand \bE         {{\bf E}}

\newcommand \bhcde      {\textsc{bhcde}}

\newcommand \bJ         {{\bf J}}
\newcommand \bJhat      {\hat{\bf J}}

\newcommand \bxhat      {\hat{\bf x}}
\newcommand \byhat      {\hat{\bf y}}
\newcommand \bzhat      {\hat{\bf z}}

\newcommand \beq        {\begin{equation}}
\newcommand \beqa	{\begin{eqnarray}}

\newcommand \cdetwo     {\textsc{cde2}}
\newcommand \cm         {\,{\rm cm}}

\newcommand \eeq	{\end{equation}}
\newcommand \eeqa	{\end{eqnarray}}

\newcommand \ercde      {\textsc{ercde}}

\newcommand \gtsim	{\gtrsim}		 

\newcommand \Lmin       {L_{\rm min}}

\newcommand \ltsim	{\lesssim}		 

\newcommand \aeff       {a_{\rm eff}}

\newcommand \Cabs       {C_{\rm abs}}
\newcommand \Cran       {C_{\rm ran}}
\newcommand \Cpol       {C_{\rm pol}}
\newcommand \Cpha       {C_{\rm pha}}
\newcommand \pol        {{\rm pol}}

\newcommand{\oldtext}[1]{}

\countdef\decade=200
\advance\decade by \year
\countdef\hours=201
\advance\hours by \time
\divide\hours by 60
\countdef\mins=202
\advance\mins by \hours
\multiply\mins by 60
\multiply\hours by 100
\countdef\miltime=203
\advance\miltime by \hours
\advance\miltime by \time
\advance\miltime by -\mins
\renewcommand\today{\number\decade.\number\month.\number\day.\number\miltime}
\begin{document}

\title{%
        {\bf On the Shapes of Interstellar Grains:
        Modeling Infrared Extinction and Polarization by Spheroids and
        Continuous Distributions of Ellipsoids}
	}

\author[0000-0002-0846-936X]{B. T. Draine}
\affiliation{Department of Astrophysical Sciences,
  Princeton University, Princeton, NJ 08544-1001, USA}
\author[0000-0001-7449-4638]{Brandon S. Hensley}
\affiliation{Department of Astrophysical Sciences,
  Princeton University, Princeton, NJ 08544-1001, USA}
\affiliation{Spitzer Fellow}

\correspondingauthor{B. T. Draine}
\email{draine@astro.princeton.edu, bhensley@astro.princeton.edu}

\begin{abstract}
Although interstellar grains are known to be aspherical, their actual shapes
remain poorly constrained.  
We assess whether three continuous distributions
of ellipsoids (CDEs) from the literature 
are suitable for describing the shapes of interstellar grains.
Randomly-selected shapes
from each distribution are shown as illustrations.
The often-used Bohren-Huffman CDE includes a very large fraction
of extreme shapes: fully 10\% of random draws have axial ratio
$a_3/a_1>19.7$, and 5\% have $a_3/a_1>33$.
The CDE2 distribution includes a much smaller fraction of extreme shapes,
and appears to be the most realistic.
For each of the three CDEs considered, we derive 
shape-averaged cross sections for extinction and polarization
in the Rayleigh limit.
Finally, we describe a method for ``synthesizing'' a dielectric
function for an assumed shape or shape distribution if the actual
absorption cross sections per grain volume in the Rayleigh limit 
are known from
observations.  This synthetic dielectric function predicts the
wavelength dependence of polarization, which can then be compared to
observations to constrain the grain shape.

\end{abstract}

\let\svthefootnote\thefootnote
\let\thefootnote\relax\footnote{\textcopyright 2020.  All rights reserved.}
\let\thefootnote\svthefootnote

  
\section{Introduction
         \label{sec:intro}}
After many years of study,
both the composition and the geometry (shape, porosity)
of interstellar grains remain uncertain.
While meteorites can provide samples of presolar grains that were
part of the interstellar grain population at the time of formation of
the solar system, the surviving particles may not be representative,
and the sampling techniques are biased toward large 
``stardust'' grains with isotopic anomalies.
Interstellar grains collide with interplanetary spacecraft, providing
some information on elemental composition, but the data are limited
and generally involve vaporization of the impinging particle, leaving
both mineralogy and preimpact morphology
uncertain \citep[e.g.,][]{Altobelli+Postberg+Fiege+etal_2016}.
The Stardust mission captured some particles relatively intact
\citep{Westphal+Bechtel+Brenker+etal_2014,
Westphal+Stroud+Bechtel+etal_2014}, but
dynamical considerations argue against these particles having come
from the interstellar medium \citep{Silsbee+Draine_2016}.

As a result, our knowledge of interstellar grains is based almost entirely
on (1) evidence of elements that have been ``depleted'' from
interstellar gas and incorporated into dust grains, and (2) observations
of the interaction of electromagnetic waves with the interstellar grains --
absorption, scattering, and emission \citep{Hensley+Draine_2021a}.
The challenge to grain modelers is to create physical models 
that are consistent with these constraints.

Grain models must specify the optical properties of the grain materials, and
the shapes and sizes of the grains.
The optical properties of a grain, particularly for polarization, depend on
the grain shape, i.e., morphology.  
Because the universe of possible grain morphologies is unbounded, 
modelers are forced to 
limit consideration to some subset of idealized shapes.
With stringent constraints now available for polarized extinction by
and emission from interstellar grains,
the assumption of spherical grains is no longer adequate for modeling.
The natural first step beyond spheres is to consider spheroids and ellipsoids.

The present work has two aims. 
The first is to discuss certain distributions of ellipsoidal shapes. 
Continuous distributions of spheroidal or ellipsoidal shapes have
been considered in some previous studies,
but the discussions have generally been limited to the 
\added{angle-averaged}
absorption cross sections, with little said about the actual 
distribution of {\it shapes}.
Here we explicitly discuss the distribution of shapes associated with
three particular continuous distributions of ellipsoids (CDEs).
We also derive the polarization cross sections for the CDEs for grains in
the ``electric dipole'' or Rayleigh limit
when the grains are not randomly oriented.

The second aim is to present a method for using observational constraints on
absorption at long wavelengths, plus a prior estimate of the
dielectric function at shorter wavelengths, to derive the complex
dielectric function $\epsilon(\lambda)$ at long wavelengths $\lambda$.
Absorption and polarization by grains both depend on the grain shape, or
distribution of shapes.  
If we knew the dielectric function $\epsilon(\lambda)$,
we could (at least in principle)
infer the actual grain shape by computing absorption 
vs.\ $\lambda$
for different assumed shape distributions, and seeing which shape
distribution best agrees with observations.  Because the actual grain
materials remain unknown, we don't know $\epsilon(\lambda)$, and hence
cannot use that approach to deduce the grain shape.
However, if we have observations of both absorption {\it and}
polarization, we can determine which shape distribution yields a dielectric
function that is consistent with both.
We show here how this can be done.
The methods developed here have been employed 
to obtain a dielectric function for ``astrodust'' 
\citep{Draine+Hensley_2021a} for continuous distributions
of ellipsoids.

The paper is organized as follows:
absorption and polarization cross sections for ellipsoids in the
long wavelength (Rayleigh) limit
are reviewed in section \ref{sec:electric dipole limit}.
In section \ref{sec:CDE} we discuss the properties of three
continuous distributions of ellipsoidal shapes -- the BHCDE, ERCDE,
and CDE2 distributions -- and
present images of shapes drawn randomly from each of these
distributions.
Analytic results for polarized absorption cross sections are presented
in sections \ref{sec:pol by CDEs} and
\ref{sec:Cabs}.
Attenuation and polarization by a medium with partial
grain alignment is discussed in section \ref{sec:polarized absorption}.
In section \ref{sec:dielectric function} we develop a method for employing
the results
obtained here, together with other constraints, to
obtain a self-consistent dielectric function given observations of
absorption as a function of wavelength.
Our results are summarized in section \ref{sec:summary}.
Certain technical results are collected in Appendices 
\ref{app:CDE}--\ref{app:one-to-one mapping}.

\section{\label{sec:electric dipole limit}
         Absorption in the Rayleigh Limit}

In the Rayleigh limit (grain size $\ll$ wavelength $\lambda$), 
the interaction of a
grain with an incident electromagnetic wave is fully characterized
by the grain's electric polarizability tensor
\citep[see, e.g.,][]{Draine+Lee_1984}.  Here we review the dependence
of this polarizability tensor on the grain shape.

\subsection{Ellipsoidal Grains}
Consider 
an ellipsoidal grain with semimajor axes $a_1\leq a_2\leq a_3$ and volume
$V=(4\pi/3)a_1a_2a_3$.
Let $\bahat_1,\bahat_2,\bahat_3$ be unit vectors along the
three principal axes.
We define an effective radius $\aeff\equiv(3V/4\pi)^{1/3}=(a_1a_2a_3)^{1/3}$.

The grain material is assumed to have an isotropic complex dielectric function
$\epsilon(\lambda)=\epsilon_1+i\epsilon_2$, where
$\epsilon_1(\lambda)$ and $\epsilon_2(\lambda)$ 
are the real and imaginary parts of $\epsilon$,
and $\lambda$ is the wavelength {\it in vacuo}.
In the long-wavelength limit $a_3\ll \lambda$,
the electric polarizability tensor for radiation with
$\bE\parallel \bahat_j$ is
$\alpha_{jj}=A_j V/4\pi$, 
where 
\beq \label{eq:A_el,j}
A_{j}(\epsilon) =
\frac{\epsilon -1}{1+L_j(\epsilon-1)}
~,
\eeq
with $L_j$ given by
\citep[see, e.g.,][]{Bohren+Huffman_1983}
\beqa \label{eq:L_j from a_j}
L_j&\,=\,&\frac{1}{2}\int_0^\infty 
\frac{dx}{\left[y_j^2+x\right]\left[(y_1^2+x)(y_2^2+x)(y_3^2+x)\right]^{1/2}}
\\ \label{eq:y_j}
y_j&\equiv& \frac{a_j}{(a_1a_2a_3)^{1/3}}
~~~.
\eeqa
The $L_j$, referred to
variously as ``geometrical factors,'' ``shape factors,'' or
``depolarization factors,'' are determined by
the axial ratios $a_1/a_3$ and $a_2/a_3$.
The $L_j$ satisfy
\beq
L_1+L_2+L_3=1
~~~.
\eeq
If $a_1\leq a_2\leq a_3$, then
\beq
L_1 \geq L_2 \geq L_3
~~~.
\eeq
The absorption cross section for radiation with
$\bE\parallel\bahat_j$ is \deleted{simply}
\beqa \label{eq:C_abs}
C_{{\rm abs},j} &\,=\,& \frac{2\pi V}{\lambda} {\rm Im}(A_j)
~~~.
\eeqa
After propagating a distance $z$ through a medium with dust number
density $n_d$, 
a plane wave will undergo both
attenuation (due to absorption) and a phase shift relative to
propagation {\it in vacuo}.  The phase shift (in radians)
will be $n_d C_{\rm pha} z$, where
\beq
C_{{\rm pha},j} = \frac{\pi V}{\lambda} {\rm Re}(A_j)
~~~.
\eeq


The axes $\bahat_1,\bahat_2,\bahat_3$ coincide with the principal axes
of the moment of inertia tensor, with eigenvalues
$I_1\geq I_2\geq I_3$.
For randomly-oriented grains the absorption cross section is
\beqa \label{eq:Cran_ed}
\Cran &\,=\,& \frac{C_\parallel+2C_\perp}{3} =
\frac{2\pi V}{\lambda}{\rm Im}\left(\frac{A_1+A_2+A_3}{3}\right)
~~~.
\eeqa
Interstellar grains are generally spinning rapidly, and it is appropriate
to average over the grain orientations.
The direction of the grain
axis $\bahat_1$ may be correlated
with the angular momentum vector $\bJ$; if the
grains are in suprathermal rotation, $\bahat_1$ will tend to be
aligned with $\bJ$, as originally pointed out by \citet{Purcell_1979}.
The absorption cross sections for $\bE \parallel \bahat_1$
and $\bE \perp \bahat_1$ are
\beqa
C_\parallel&\,\equiv\,&C_{\rm abs}({\bf E}\parallel\bahat_1)
=
\frac{2\pi V}{\lambda} {\rm Im}(A_1)
\\
C_\perp&\equiv&C_{\rm abs}({\bf E}\perp\bahat_1)
=
\frac{2\pi V}{\lambda} {\rm Im}\left(\frac{A_2+A_3}{2}\right)
~~~,
\eeqa
where the grains are assumed to be spinning with $\bahat_2$
and $\bahat_3$ randomly-distributed in the plane $\perp$ to $\bahat_1$.

Consider the limiting case of
spinning grains that are perfectly-aligned
with $\bahat_1\parallel\bJ$.
For unpolarized radiation
propagating with wavevector
${\bf k}\perp\bJ$, the polarization-averaged absorption
cross section is
\beq
C_{\rm abs}= \frac{C_\perp+C_\parallel}{2}
= \frac{2\pi V}{\lambda} {\rm Im}\left(\frac{A_2+A_3+2A_1}{4}\right)
~~~.
\eeq
The difference in absorption cross
sections will produce linear polarization, characterized by
the ``polarization cross section''
\beq \label{eq:Cpol}
\Cpol = \frac{C_\perp-C_\parallel}{2} =
\frac{2\pi V}{\lambda}
{\rm Im}\left(\frac{A_2+A_3-2A_1}{4}\right)
~~~.
\eeq
There will also be a phase shift between the two linear polarizations.
We define
\beq
\Delta C_{\rm pha} \equiv C_{{\rm pha},\perp}-C_{{\rm pha},\parallel}
= \frac{\pi V}{\lambda} {\rm Re}\left(\frac{A_2+A_3-2A_1}{2}\right)
~~~.
\eeq
After propagating a distance $z$ through a medium with dust number density
$n_d$, the phase difference
between the modes will be $n_d\Delta C_{\rm pha} z$.
If the direction of grain alignment rotates along the direction
of propagation, radiation that is initially unpolarized will
develop circular polarization \citep{Martin_1972b,Martin_1974}.
We define a ``circular polarization efficiency factor''
\beq
Q_{\rm cpol}\equiv \frac{\Cpol}{\pi \aeff^2}\times
\frac{ \Delta\Cpha}{\pi \aeff^2}
~~~.
\eeq
If the rotation angle is small, and the percentage linear polarization
is small, the circular polarization after propagating a pathlength $z$ 
has Stokes parameters ${\rm V}$ and ${\rm I}$ varying as
\beq
\frac{\rm V}{\rm I} \propto Q_{\rm cpol} \times (n_d \pi \aeff^2 z)^2
~~~.
\eeq
\subsection{Spheroids}

Prolate spheroids have $a_1=a_2 < a_3$, and oblate spheroids
have $a_1<a_2=a_3$.
The ``shape factors'' $L_j$ are given by
\citep{van_de_Hulst_1957}
\beqa
{\rm prolate:}&&L_3 = \frac{1-e^2}{e^2}
\left[\frac{1}{2e}\ln\left(\frac{1+e}{1-e}\right)-1\right] < 1/3
\hspace*{0.5em},\hspace*{0.5em}
e^2\equiv1-\left(\frac{a_1}{a_3}\right)^2
\\
&&L_1=L_2=\frac{1-L_3}{2}
\\
{\rm oblate:}&&L_1 = \frac{1+e^2}{e^2}
\left[1-\frac{1}{e}\arctan(e)\right] > 1/3
\hspace*{0.5em},\hspace*{0.5em}
e^2\equiv\left(\frac{a_3}{a_1}\right)^2 -1
\\
&&L_2=L_3=\frac{1-L_1}{2}
~~~.
\eeqa
A sphere has $(L_1,L_2,L_3)=(\frac{1}{3},\frac{1}{3},\frac{1}{3})$;
the prolate limit (needle-like) has 
$(L_1,L_2,L_3)=(\frac{1}{2},\frac{1}{2},0)$;
the oblate limit (disk-like) has
$(L_1,L_2,L_3)=(1,0,0)$.

\section{\label{sec:CDE}
                   Continuous Distributions of Ellipsoids}
\subsection{Shape Factors}
\begin{figure}[ht]
\begin{center}
\includegraphics[angle=0,width=10.0cm,
                 clip=true,trim=0.5cm 5.0cm 0.5cm 2.5cm]
                {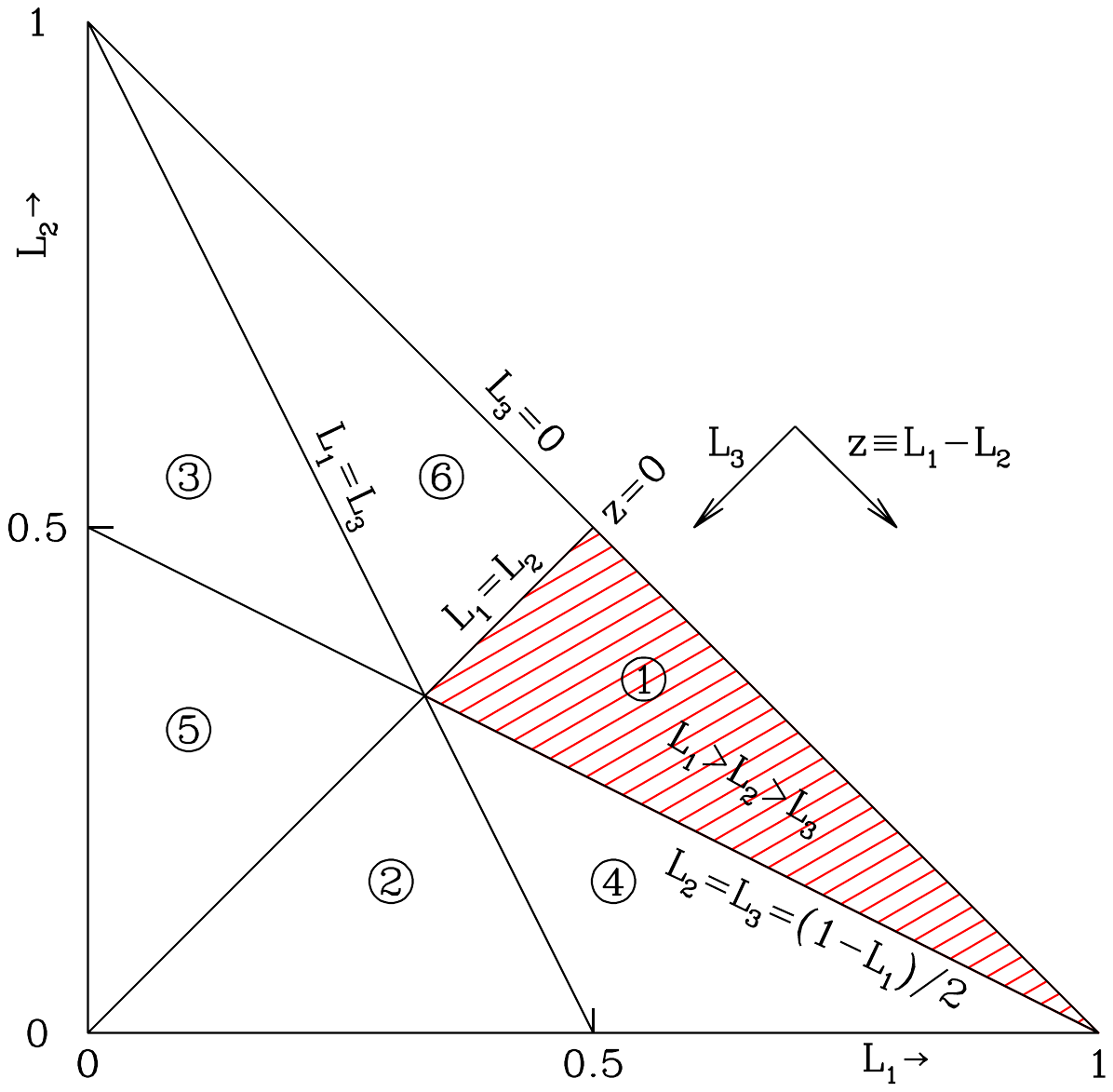}
\caption{\label{fig:triangle}
         The domain of allowed shape factors $(L_1,L_2)$.
         The shaded region is the domain where 
         $L_3 \leq L_2 \leq L_1$ (see text).
         Other regions, numbered 2-6, correspond to the other
         possible orderings of $L_1,L_2,L_3$.}
\end{center}
\end{figure}

Every ellipsoidal shape is uniquely specified by its triplet of
depolarization factors $(L_1,L_2,L_3)$.
Consider a population of ellipsoidal grains, each with the same volume $V$,
but with some continuous distribution of axial ratios -- this is
referred to as a ``continuous distribution of ellipsoids'' (CDE).
Suppose that each grain has principal axes labelled 1,2,3 
arbitrarily, and
that $G(\ell_1,\ell_2)d\ell_1 d\ell_2$ is the fraction of
the population with $L_1\in[\ell_1,\ell_1+d\ell_1]$,
$L_2\in[\ell_2,\ell_2+d\ell_2]$, and $L_3=1-L_1-L_2$.
The function $G$ is non-negative ($G\geq 0$) and normalized: 
$\int G(L_1,L_2)dL_1dL_2=1$ over the allowed
$(L_1,L_2)$ domain.
If labels 1,2,3 were assigned arbitrarily, 
the function $G$ must satisfy symmetry requirements, including
$G(L_1,L_2)=G(L_2,L_1)=G(L_1,1-L_1-L_2)$,\footnote{
    One can also consider functions $G$ that do not satisfy these
    symmetry requirements, but in this case one must restrict discussion
    to only one of the six subregions in Figure \ref{fig:triangle}.}
but otherwise
we have no a-priori knowledge of the function $G$, other than
expecting that very extreme axial ratios should be rare.

\begin{figure}[h]
\begin{center}
\includegraphics[angle=0,width=17.0cm,clip=true,
                 trim=0.1cm 0.0cm 0.0cm 0.0cm]
{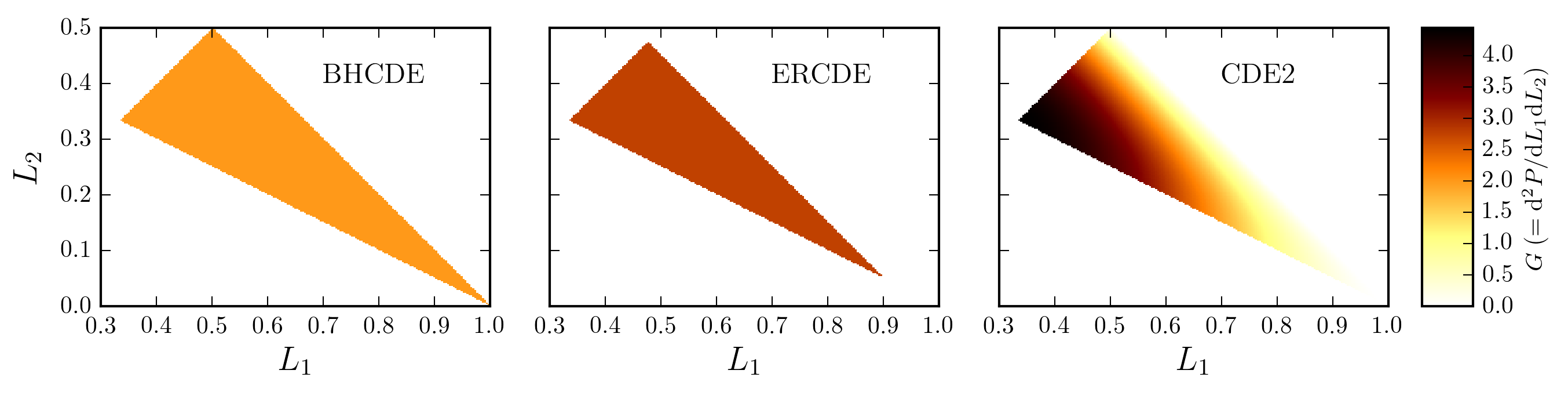}
\caption{\label{fig:shape_factors}\footnotesize 
   $G(L_1,L_2)$ for the BHCDE, ERCDE (with $\Lmin=0.05$) and
   CDE2 shape distributions.}
\end{center}
\end{figure}

Various distributions of shapes have been considered in
the literature, including
spheroids 
\citep{Treffers+Cohen_1974,Min+Hovenier+deKoter_2003},
and ellipsoids 
\citep{Bohren+Huffman_1983}.
\citet{Bohren+Huffman_1983} gave a lucid introduction to CDEs in general,
and presented a simple illustrative example, referred to here
as the BHCDE.
We discuss the BHCDE and two other distributions of ellipsoids 
that have been considered in the astrophysical literature.
\begin{enumerate}
\item {\bf BHCDE:\,} The simplest functional form
\beq \label{eq:BHCDE}
G(L_1,L_2)=2
~~~{\rm for}~L_1\geq0, ~ L_2\geq 0, ~ L_1+L_2\leq 1
\eeq
is often considered; \citet{Bohren+Huffman_1983} present this
as an example, and it has subsequently been applied
by a number of authors
\citep[e.g.,][]{Rouleau+Martin_1991,
Alexander+Ferguson_1994,Min+Hovenier+deKoter_2003,
Min+Hovenier+Dominik+etal_2006,
Sargent+Forrest+DAlessio+etal_2006,
Min+Hovenier+Waters+deKoter_2008,
Rho+Gomez+Boogert+etal_2018}.
Because $G(L_1,L_2)$ is independent of $L_1$ and $L_2$, it is sometimes
asserted that ``all shapes are equally probable''
\citep{Bohren+Huffman_1983} or ``all shapes are equally weighted''
\citep{Sargent+Forrest+DAlessio+etal_2006}, seemingly
suggesting that this is a ``fair'' sampling of ellipsoidal shapes.
While it is correct that all ellipsoidal shapes are present, 
it is not clear how ``all shapes are equally probable'' 
is to be understood,
given that shapes are not discrete and
there is no commonly accepted metric for ``shape space''.

Although having the virtue of analytic simplicity,
we will see below that the BHCDE distribution has an extreme
representation of very elongated shapes, with $L\rightarrow 0$.
We will argue that the BHCDE distribution seems unlikely to approximate
grain shape distributions in nature, whether for desert sand or 
interstellar dust.

\item {\bf ERCDE:\,} 
\citet{Zubko+Mennella+Colangeli+Bussoletti_1996} proposed eliminating
the most extreme shapes by truncating
the distribution (\ref{eq:BHCDE}):
\beq \label{eq:ERCDE}
G(L_1,L_2)=\frac{2}{(1-3\Lmin)^2}
~~~{\rm for}~ L_1\geq\Lmin, ~ L_2\geq\Lmin, ~ L_1+L_2 \leq 1-\Lmin
~~~,
\eeq
referring to this as the ``externally-restricted CDE'' (ERCDE).
$\Lmin$ is a free parameter.
While removing extreme shapes with $L_j\rightarrow 0$ 
or $L_j\rightarrow 1$
is desirable, 
the ERCDE distribution still seems unphysical, as we will see below.
Note that if $\Lmin\rightarrow0$, the ERCDE\,$\rightarrow$\,BHCDE.

\item {\bf CDE2:\,} \citet{Ossenkopf+Henning+Mathis_1992} proposed
the distribution
\beq \label{eq:CDE2}
G(L_1,L_2)=120 L_1L_2L_3=120 L_1L_2(1-L_1-L_2)
~~~{\rm for}~L_1\geq0, ~ L_2\geq 0, ~ L_1+L_2\leq 1
~~~,
\eeq
which has the desirable behavior $G\rightarrow 0$ for $L_3\rightarrow 0$
and $L_1\rightarrow 1$.
This distribution has subsequently been referred to as
``CDE2''
\citep{Fabian+Henning+Jager+etal_2001,Sargent+Forrest+DAlessio+etal_2006},
and we shall so refer to it here.
\end{enumerate}
The distribution functions $G(L_1,L_2)$ for these three CDEs are
shown in Figure \ref{fig:shape_factors}.

\subsection{\label{sec:CDE shapes}
            Shape Distributions}

Because the optical properties of ellipsoids in the limit
$a\ll\lambda$ are determined by
$L_1$, $L_2$, and $L_3=1-L_1-L_2$, 
most discussions of CDEs
have been concerned only with the distribution of $L_j$ values,
rather than the distributions of the ellipsoid axial ratios.
However, it is of interest to
examine the distributions of actual grain shapes that correspond to
the BHCDE, ERCDE, and CDE2 distributions.

For a given set of axial ratios $(a_2/a_1,a_3/a_1)$, the
$L_j$ values can be obtained by numerical quadrature
[Eq.\ (\ref{eq:L_j from a_j})].
Since there does not appear to be any direct way 
to invert Eq.\ (\ref{eq:L_j from a_j}) to obtain $(a_2/a_1,a_3/a_1)$
from given $(L_1,L_2)$, we have implemented a numerical procedure to
find $(a_2/a_1,a_3/a_1)$ corresponding to given $(L_1,L_2)$.
In Appendix \ref{app:one-to-one
mapping}, we demonstrate that any solution found in this way is unique.

\begin{figure}[h]
\begin{center}
\includegraphics[angle=0,width=12.0cm,clip=true,
                 trim=0.5cm 5.0cm 0.5cm 2.5cm]
{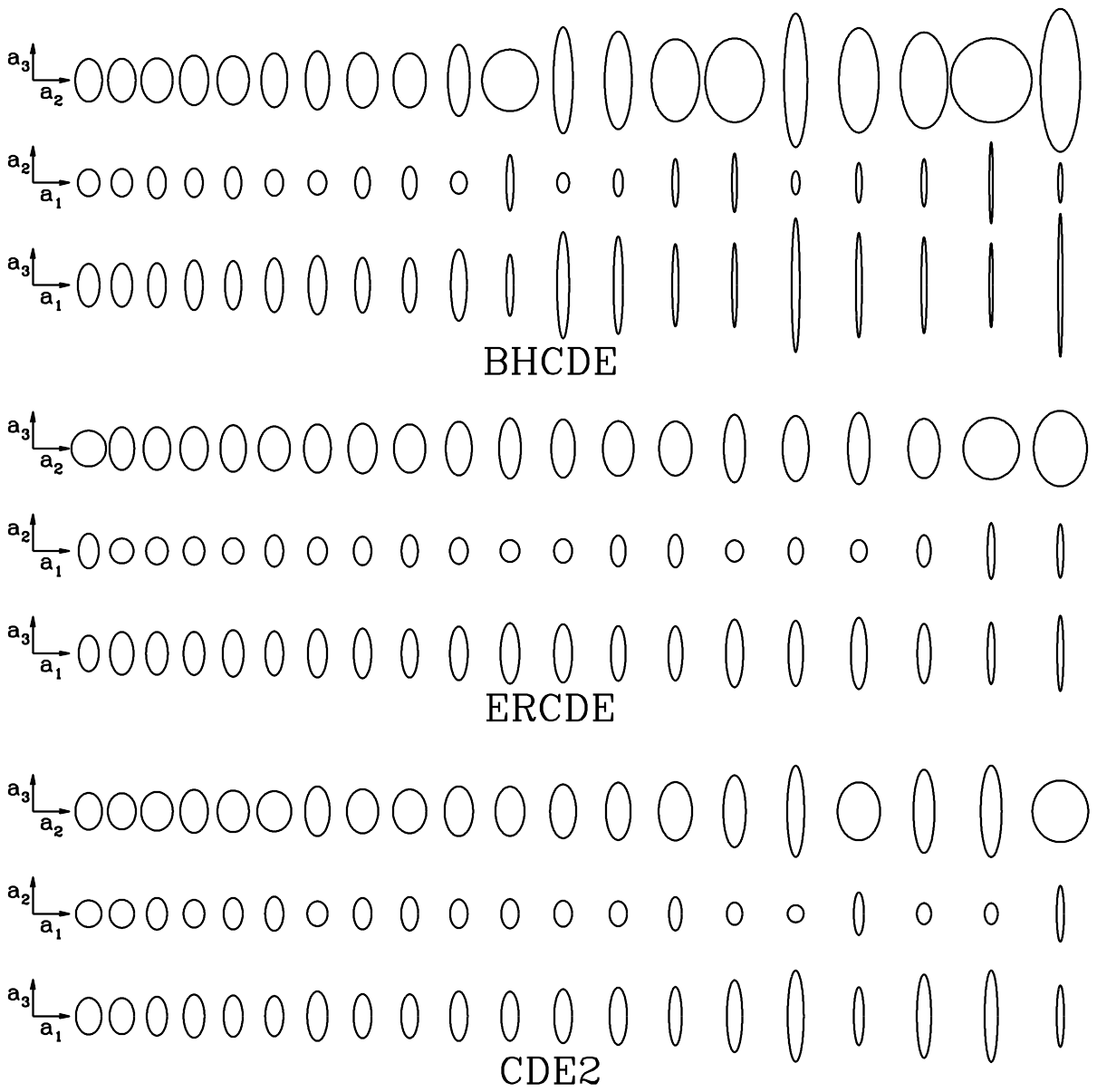}
\caption{\label{fig:shapes}\footnotesize
   20 randomly-selected ellipsoids drawn 
   from the BHCDE, ERCDE, and CDE2 distributions.
   All examples have equal volume.  3 views are shown for each shape:
   viewed along the short axis $\bahat_1$ (top row),
   and along the $\bahat_3$ and $\bahat_2$ axes (2nd and 3rd rows).
   For each distribution the 20 random 
   shapes are shown in order of increasing $a_3/a_1$ (left to right).}
\end{center}
\end{figure}

\begin{figure}[h]
\begin{center}
\includegraphics[angle=0,width=8.0cm,clip=true,
                 trim=0.5cm 5.0cm 0.5cm 2.5cm]
{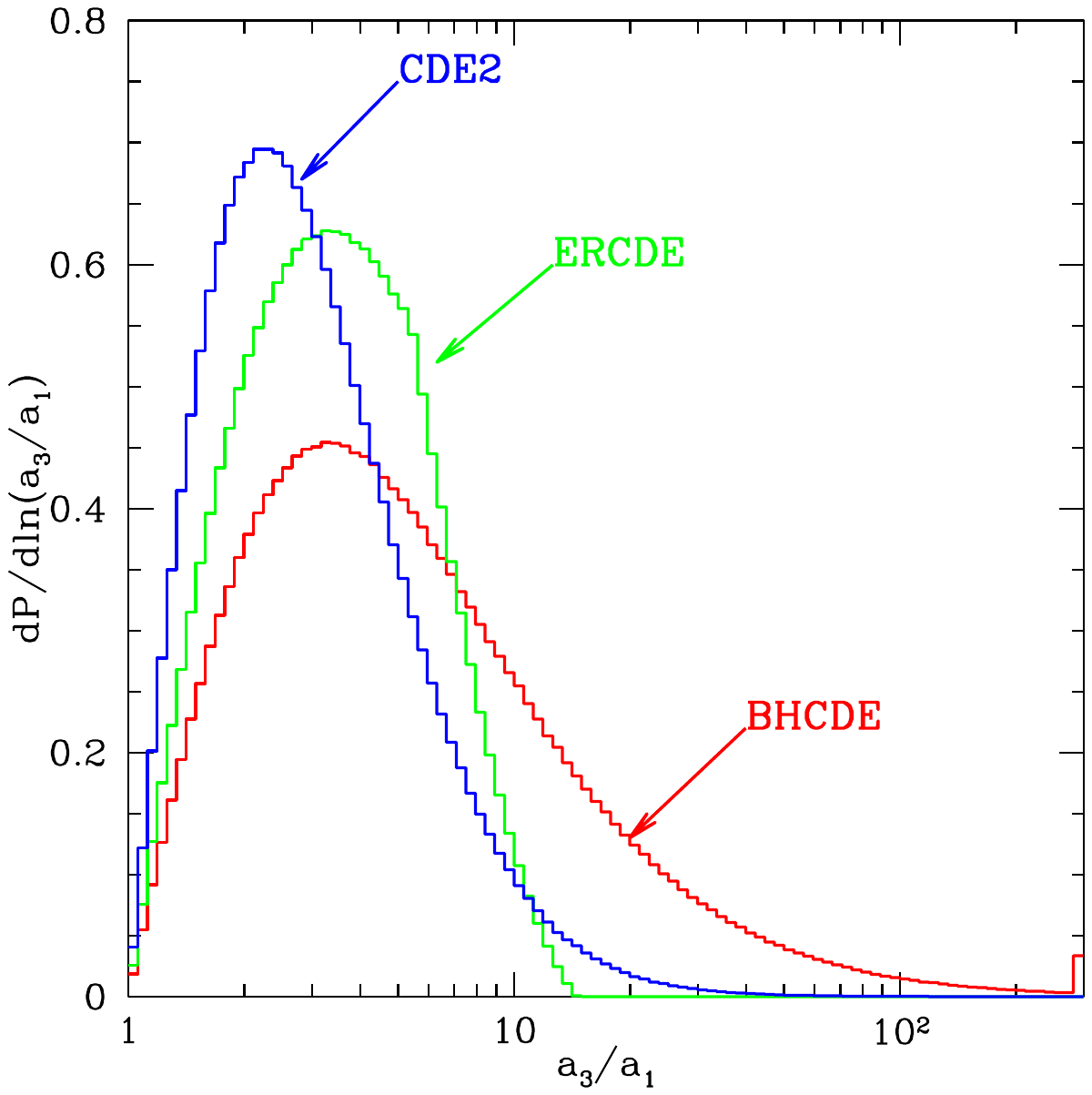}
\includegraphics[angle=0,width=8.0cm,clip=true,
                 trim=0.5cm 5.0cm 0.5cm 2.5cm]
{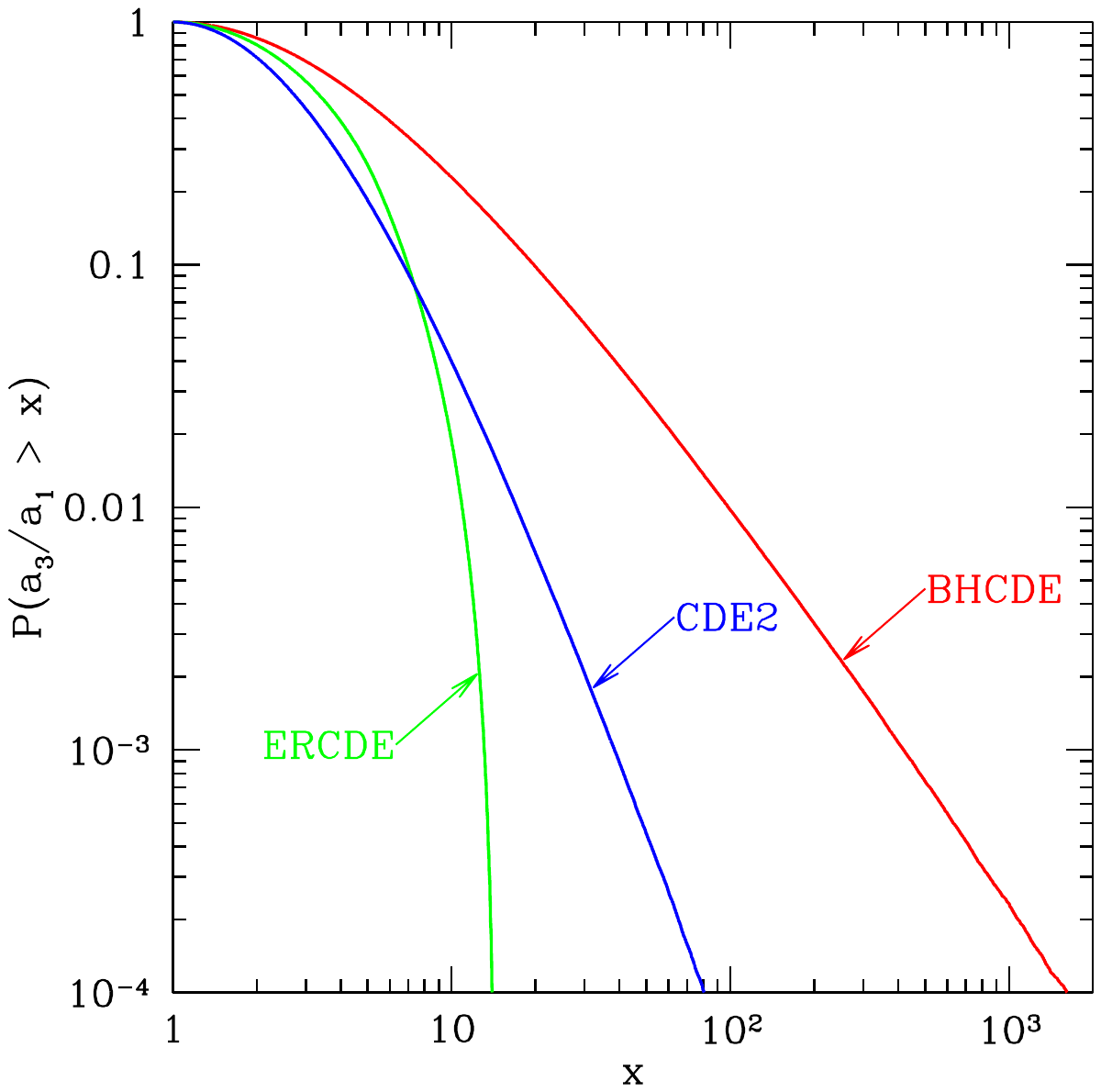}
\caption{\label{fig:hist}\footnotesize
   (a) Distribution of long/short axial ratio $a_3/a_1$ for three continuous
   distributions of ellipsoids.  The ERCDE with $L_{\min}=0.05$ has
   a maximum allowed axial ratio $a_3/a_1=14$, but the CDE2 and BHCDE
   distributions both extend to infinite axial ratios.  The
   BHCDE distribution has a much larger representation of extreme
   axial ratios.
   (b) Cumulative distribution functions.
   For the BHCDE distribution, 10\% of the realizations have
   $a_3/a_1>19.7$, and 1\% have $a_3/a_1>98.5$.}
\end{center}
\end{figure}


\begin{figure}[h]
\begin{center}
\includegraphics[angle=0,width=17.0cm,
                  clip=true,trim=0.3cm 0.0cm 0.0cm 0.0cm]
{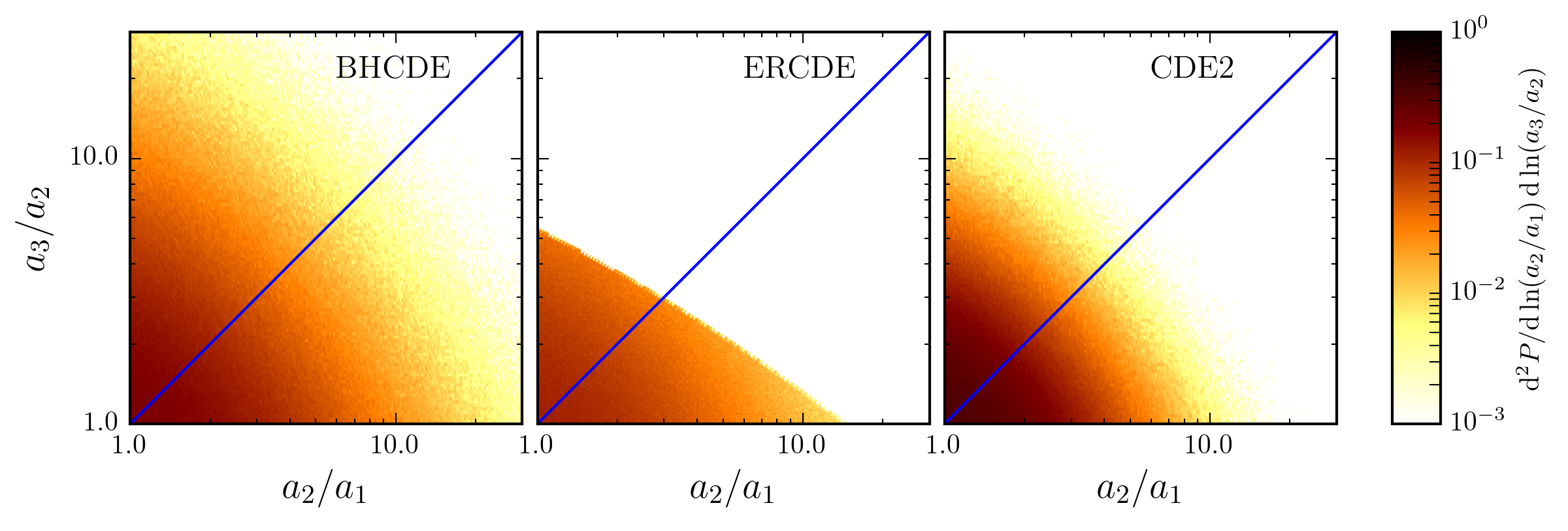}
\caption{\label{fig:axrats}\footnotesize
   Distributions of axial ratios $a_3/a_2$ and $a_2/a_1$ 
   for the BHCDE, ERCDE ($\Lmin=0.05$), 
   and CDE2 shape distributions.
   Oblate spheroids have $a_3/a_2=1$, and prolate
   spheroids have $a_2/a_1=1$.}
\end{center}
\end{figure}

\begin{table}[b]
\begin{center}
\caption{\label{tab:hist and cdf}
                Long/Short Axis Ratio $a_3/a_1$}

\begin{tabular}{|cccc|}
\hline
        & BHCDE & ERCDE$^b$ & CDE2 \\
\hline
mode$^a$ & 3.26   & 3.27   & 2.24 \\
median   & 4.58   & 3.35   & 2.73 \\
25\%     & 9.23   & 5.07   & 4.25 \\
10\%     & 19.7   & 6.97   & 6.72 \\
5\%      & 32.97  & 8.32   & 9.11 \\
1\%      & 98.49  & 10.92  & 17.11 \\
\hline
\multicolumn{4}{l}{$a$~~ Maximum of $dP/d\ln(a_3/a_1)$.}\\
\multicolumn{4}{l}{$b$~~ $L_{\rm min}=0.05$.}\\
\end{tabular}
\end{center}
\end{table}
We continue to adopt the ordering $a_1\leq a_2\leq a_3$,
$L_1\geq L_2 \geq L_3$.
We draw $(L_1,L_2)$ values randomly according to the BHCDE, ERCDE, or CDE2
distributions, and for each $(L_1,L_2)$ find the corresponding axial ratios
$(a_2/a_1,a_3/a_1)$.
Figure \ref{fig:shapes} shows 20 examples selected randomly from
each of these shape distributions.
Figure \ref{fig:hist}a shows the distribution of long/short axial ratios
$a_3/a_1$ for the BHCDE, ERCDE (with $L_{\min}=0.05$), and CDE2
distributions.
Figure \ref{fig:hist}b shows the cumulative distribution function of
axial ratios $a_3/a_1$, and
Figure \ref{fig:axrats} shows the distributions of axial ratios for the
BHCDE, ERCDE, and CDE2 distributions.
Some characteristics of these shape distributions are listed in Table 
\ref{tab:hist and cdf}.

The BHCDE distribution has a very large fraction of extreme
axial ratios -- Figure \ref{fig:hist}b shows that
10\% of the realizations have
$a_3/a_1>19.7$, and 1\% of the realizations have $a_3/a_1>98.5$.
\added{Extreme elongation will increase the susceptibility to fragmentation
in high-speed grain-grain collisions.  Highly elongated grains may
also be more vulnerable to centrifugal disruption
if spun-up by strong radiative
torques \citep{Silsbee+Draine_2016,Hoang_2019}
or gas-grain streaming \citep[e.g.,][]{Tatsuuma+Kataoka_2021}.} 
The actual shape distribution for interstellar grains is of course
unknown, but it seems unlikely to include as large
a fraction of extreme aspect ratios as
the BHCDE distribution.
The CDE2 (with $\sim$90\% of the draws having
$a_3/a_1<6.72$) or ERCDE (with $\sim$90\% of the draws having
$a_3/a_1<6.9$ for $\Lmin=0.05$)
may be more plausible shape distributions to consider
for interstellar dust grains.

\section{\label{sec:pol by CDEs}
         Polarization by CDEs}

The observed polarization of starlight by dust, 
and of submm emission from dust
\added{in the interstellar medium}, 
indicates that interstellar grains spin with their short
axis tending to be aligned with the local magnetic field $\bB$; this occurs
because the grain's angular momentum $\bJ$
tends to
align with the magnetic field, and
the short axis of the grain tends to align with $\bJ$.
Rotation and nutation, and precession of $\bJ$ around
$\bB$, are all rapid,
\added{and physical processes such as paramagnetic dissipation cause
$\bJ$ to align with $\bB$. 

In protoplanetary disks, magnetic effects
are relatively much weaker.  Grain drift can cause $\bJ$ to tend to
be perpendicular to the (azimuthal) streaming direction \citep{Gold_1952}, 
while radiative torques may cause $\bJ$ to tend toward the radial direction
\citep{Lazarian+Hoang_2007a,Tazaki+Lazarian+Nomura_2017}.
Whatever the spin-up process, if the grains are spinning suprathermally
we expect dissipation in the grain to cause
the short axis to be aligned with $\bJ$.  
The results obtained below for absorption
cross sections averaged over CDEs
are applicable both to the interstellar medium and to protoplanetary disks.
This is true also for the polarization cross sections,
provided only that the degree of alignment of the short axis with $\bJ$
is independent of shape.
Interpretation of observed polarization is often complicated by the
need to include polarized scattering, which can even be
important at submm wavelengths in protoplanetary disks
\citep{Kataoka+Muto+Momose+etal_2015}.}

In order to discuss polarization by a population of partially-aligned
grains, we require the distribution
of depolarization factors separately for the short axis, and
for the other two axes.

It is useful to restrict
consideration to the ordering $0\leq L_3\leq L_2\leq L_1\leq 1$:
for each ellipsoid,
$j=3$ corresponds to the long axis, $j=1$ to the short axis,
and $j=2$ to the intermediate axis.
Let $g_j(\ell)d\ell$ be the fraction of ellipsoids
with $L_j\in [\ell,\ell+d\ell]$.
The distribution functions $g_1,g_2,g_3$ 
can be obtained from $G$, as discussed in
Appendix \ref{app:CDE}.
Figure \ref{fig:gj} shows $g_1$, $g_2$, and $g_3$ for the BHCDE,
ERCDE, and CDE2 shape distributions.

\begin{figure}[ht]
\begin{center}
\includegraphics[width=8.0cm,angle=0,
                 clip=true,trim=0.5cm 0.5cm 0.5cm 0.5cm]
                {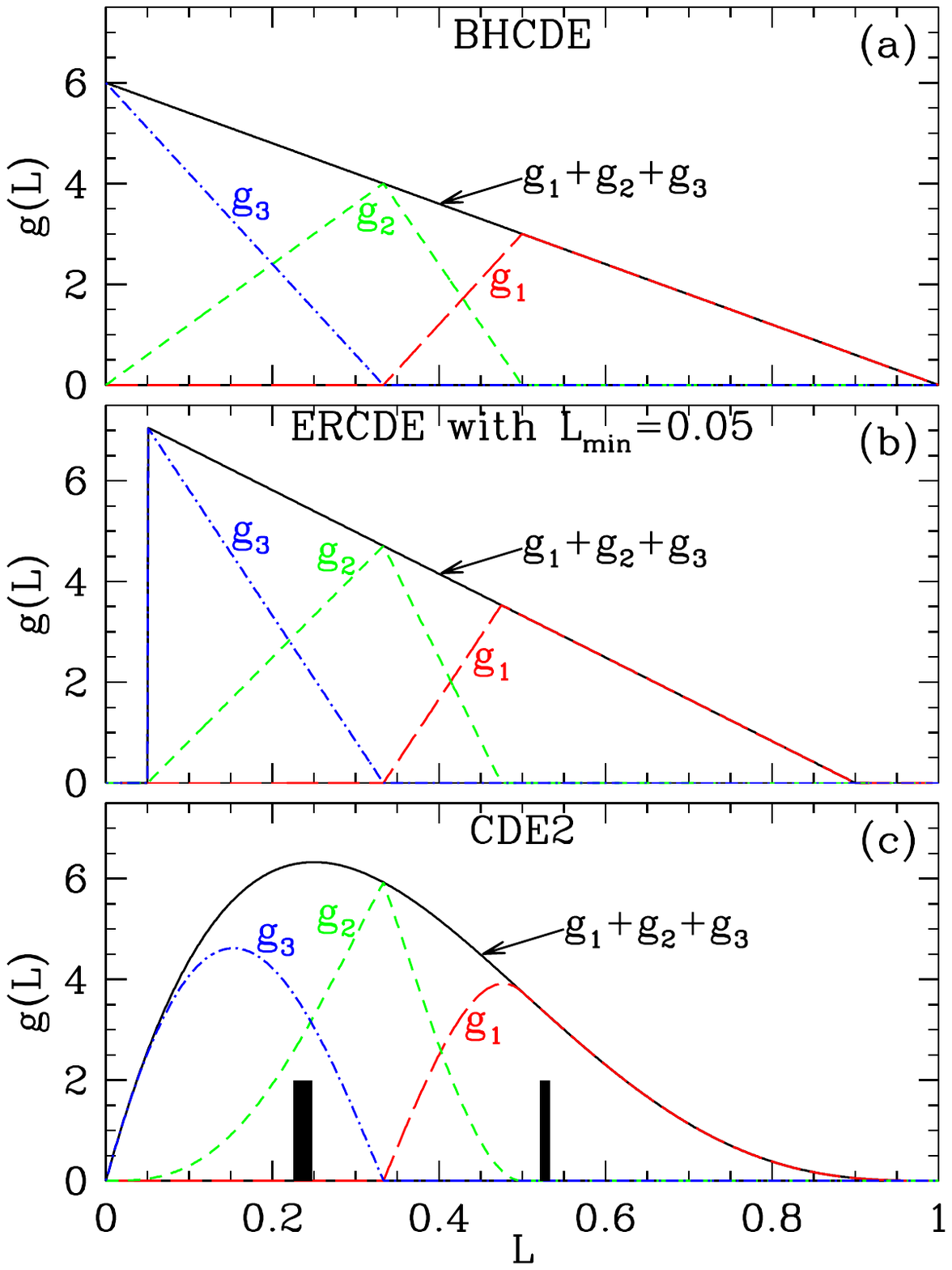}
\caption{\label{fig:gj} \footnotesize
         Distributions $g_1$, $g_2$, and $g_3$ of depolarization factors 
         $L_1,L_2,L_3$
         for the three CDE distributions discussed in \S\ref{sec:CDE}:
         (a) Bohren \& Huffman CDE (BHCDE) from Eq.\ (\ref{eq:BHCDE}).
         (b) Externally-restricted CDE (ERCDE)
         from Eq.\ (\ref{eq:ERCDE}) with $\Lmin=0.05$.
         (c) Ossenkopf, Henning \& Mathis CDE (CDE2) 
         from Eq.\ (\ref{eq:CDE2}).
         The BHCDE and ERCDE distributions have extreme representation
         of small $L$.
         Of the three,
         the CDE2 distribution appears most
         realistic (see text).
         The solid bars in panel (c) show the values of $L_3=L_2=0.2364$
         and $L_1=0.5272$ for a $b/a=2$ oblate spheroid.}
\end{center}
\end{figure}

\subsection{The BHCDE Distribution}
Figure \ref{fig:gj}a shows the distribution functions
$g_j$ for the BHCDE distribution.
We see that $g_3(\ell)$ peaks at
$\ell=0$, corresponding to infinitely elongated needles:
fully 10\% of the BHCDE ellipsoids have $L_3<0.0171$.
Only very extreme shapes have such small values of $L_3$ -- for
example, a prolate spheroid with axial ratios 1\,:\,1\,:\,11.17
has $L_3=0.0171$.
Another example with $L_3=0.0171$ would be
an ellipsoid with axial ratios 1\,:\,4.72\,:\,22.3.
It does not seem likely (to us) that interstellar grains
will have such a large fraction of extremely elongated shapes.

\subsection{The ERCDE Distribution}

The ERCDE distribution is similar to the BHCDE distribution, except
that cases with $L < \Lmin$ are excluded.  Thus $\Lmin$
is a free parameter for the ERCDE distribution.
The ERCDE distribution has $g_3$ peaking at $L_3=\Lmin$.
As an example, we consider $\Lmin=0.05$ (see Figure \ref{fig:gj}c).

What shapes would correspond to the limiting cases $L_3=\Lmin$?
One example of a shape with $L_3=0.05$:
a prolate spheroid with axial ratios 1\,:\,1\,:\,5.41 
(with $L_1=L_2=0.475$, $L_3=0.05$).
Another example: an oblate spheroid with
axial ratios 1\,:\,14.43\,:\,14.43
(with $L_1=0.9$, $L_2=L_3=0.05$).
A third example: an ellipsoid with axial ratios
1\,:\,2.965\,:\,8.79 (with $L_1=0.719$, $L_2=0.231$, $L_3=0.05$).

Because $g_3$ increases monotonically as $L_3\rightarrow\Lmin$
(see Figure \ref{fig:gj}c), this shape distribution places substantial
weight on the most extreme allowed grain shapes.
For instance, 
fully 10\% of the ERCDE realizations with $\Lmin=0.05$ have $L_3<0.06454$.
Thus the ERCDE shape distribution also appears to 
overrepresent extreme shapes, unless $\Lmin\gtsim0.10$.
The ERCDE shape distribution will be further discussed below.

\subsection{The CDE2 Distribution}

The distribution functions $g_j$ for the CDE2 distribution are shown in
Figure \ref{fig:gj}c.
While the CDE2 does include extreme shapes, it has $g_3\rightarrow 0$ for
$L_3\rightarrow 0$, and $g_1\rightarrow 0$ for $L_1\rightarrow 1$.
10\% of the realizations have $L_3<0.06185$, so it is somewhat
similar to the ERCDE with $\Lmin=0.05$ 
in the representation of extreme shapes,
although the CDE2 distribution function has the virtue of smoothness.


\section{\label{sec:Cabs}
         Absorption Cross Sections for the BHCDE, ERCDE and CDE2 Distributions}

The shape-averaged absorption cross section associated with axis $j$ is
\beqa \label{eq:Cabs and A_j}
\Cabs(\bE\parallel\bahat_j) &\,=\,& \frac{2\pi V}{\lambda}
{\rm Im}(\langle A_j\rangle)
\\ \label{eq:CDEint}
\langle A_j\rangle &\equiv&
\int A_j \, g_j(\ell) d\ell
~~~,
\eeqa
where $A_j$ is related to the complex dielectric function
$\epsilon$ through Eq.\ (\ref{eq:A_el,j}).
For ellipsoids with specified axial ratios $a_1\!:\!a_2\!:\!a_3$, 
the $g_j$ are $\delta$-functions.
As seen above, for a population of 
ellipsoids with a continuous distribution of shapes, the
$g_j$ become continuous distribution functions (see Fig.\ \ref{fig:gj}).
\citet{Min+Hovenier+Dominik+etal_2006} show that
a single particle with an irregular shape also has its
absorption cross section given by Eq.\ (\ref{eq:Cabs and A_j})
with continuous distribution functions $g_j$.

\subsection{Randomly-Oriented Particles}

For randomly-oriented particles, the absorption cross section is
\beq
\Cran = 
\frac{2\pi V}{\lambda} {\rm Im}
\left[
\frac{\langle A_1\rangle + \langle A_2\rangle + \langle A_3\rangle}{3}
\right]
~~~.
\eeq
\citet{Bohren+Huffman_1983} obtained the absorption cross section
for randomly-oriented grains with the BHCDE shape distribution:
\beqa 
\label{eq:Cran_BHCDE} 
\frac{\Cran^\bhcde}{V}
&\,=\,&
\frac{4\pi}{\lambda} {\rm Im}
\left[\left(\frac{1+x}{x}\right)\ln\epsilon\right]
~~~,
\eeqa
where $x\equiv \epsilon-1$.
For the ERCDE [Eq.\ (\ref{eq:ERCDE})]
the absorption cross section 
for randomly-oriented grains was obtained by 
\citet{Zubko+Mennella+Colangeli+Bussoletti_1996}:
\beqa \label{eq:Cran for ERCDE}
\frac{\Cran^\ercde}{V}
&\,=\,&
\frac{4\pi}{\lambda}
\frac{1}{(1-3\Lmin)^2}
{\rm Im}\left\{\left(\frac{1}{x}+D\right)
\ln\left[\frac{1+xD}{1+x\Lmin}\right]
\right\} ~~~,
\eeqa
where $D\equiv 1-2\Lmin$.
It is easily verified that this reduces to Eq.\ (\ref{eq:Cran_BHCDE})
for $\Lmin\rightarrow 0$.

\citet{Fabian+Henning+Jager+etal_2001} obtained the absorption cross section
for randomly-oriented ellipsoids with the CDE2 shape distribution 
[Eq.\ (\ref{eq:CDE2})]:
\beqa \label{eq:Cran for CDE2}
\frac{\Cran^\cdetwo}{V} 
&\,=\,& \frac{40\pi}{\lambda} {\rm Im}
\left[\frac{1}{x^4}
\left(-(1+x)^3\ln(1+x)+x+\frac{5}{2}x^2+\frac{11}{6}x^3+\frac{1}{4}x^4\right)
\right]
~~~.
\eeqa

\subsection{Polarization Cross Sections for Aligned Particles}
The polarization cross section (see Eq.\ \ref{eq:Cpol}) is
\beq
\Cpol \equiv \frac{\pi V}{\lambda}{\rm Im}
\left[
\frac{\langle A_2\rangle + \langle A_3\rangle - 2\langle A_1\rangle}{2}
\right]
~~~.
\eeq
For the BHCDE distribution, we find
\beqa \nonumber
\frac{\Cpol^\bhcde}{V} &\,=\,&
\frac{3\pi}{2\lambda}
{\rm Im}\Big[
\frac{12}{x}\left(1+\frac{x}{2}\right)\ln\left(1+\frac{x}{2}\right)
-\frac{9}{x}\left(1+\frac{x}{3}\right)\ln\left(1+\frac{x}{3}\right)
\\ \label{eq:Cpol_BHCDE}
&&\hspace*{2.0cm}
-\frac{2}{x}\left(1+x\right)\ln\left(1+x\right)
\Big]
~~~.
\eeqa
where $x\equiv\epsilon-1$.
For the ERCDE distribution we find
\beqa
\frac{\Cpol^\ercde}{V} \nonumber
&\,=\,&
\frac{3\pi}{2\lambda}
\frac{1}{(1-3\Lmin)^2}
{\rm Im}\Bigg\{
12\left(\frac{1}{x}+B\right)
\ln\left[1+xB\right]
-9\left(\frac{1}{x}+\frac{1}{3}\right)
\ln\left[1+\frac{x}{3}\right]
\\ \label{eq:Cpol_ERCDE}
&&\hspace*{9.5em}
-\left(\frac{1}{x}+D\right)
\ln\left[1+x\Lmin\right]
-2\left(\frac{1}{x}+D\right)
\ln\left[1+xD\right]
\Bigg\}
~~~~
\\
B&\equiv& \frac{1}{2}-\frac{\Lmin}{2}
\hspace*{2em},\hspace*{2em}
D\equiv 1-2\Lmin
~~~.
\eeqa
See Appendix \ref{app:CDE} for the derivation of Eq.\ (\ref{eq:Cpol_ERCDE}).
Eq.\ (\ref{eq:Cpol_BHCDE}) is recovered by setting
$L_{\min}=0$. 

The polarization cross section for the CDE2 distribution is
(see Appendix \ref{app:CDE}):
\beqa
\nonumber
\frac{\Cpol^\cdetwo}{V} 
&\,=\,& \frac{30\pi}{\lambda}
{\rm Im}\Bigg[
\frac{1}{x^4}
\Bigg(
3\left(-9-3x+3x^2+x^3\right)\ln\left(1+\frac{x}{3}\right)
+
6\left(4-3x^2-x^3\right)\ln\left(1+\frac{x}{2}\right)~~~
\\
&&
~~~~~~~~~~~~~~~~~~~~
+
2(1+x)^3\ln(1+x)
-5x-\frac{1}{2}x^2+\frac{7}{6}x^3+\frac{11}{72}x^4
\Bigg)
\Bigg]
~~~.
\eeqa

\section{\label{sec:polarized absorption}
         Polarized Absorption by Partially-Aligned Grains}


An interstellar grain with angular momentum $\bJ$ will have a magnetic
moment $\bmu$ resulting from a combination of
the Barnett effect (if the grain has unpaired electron spins),
the Rowland effect (if the grain is charged),
and ferromagnetism (if the grain contains magnetic material).\footnote{
  For ferromagnetic grains,
  the rotation-averaged effective magnetic moment
  $\langle\bmu\rangle=\bJ\langle\bJ\cdot\bmu\rangle/J^2$.}
If 
$|\bJ\times\bB|\neq0$,
the $\bmu\times\bB_0$ torque will cause $\bJ$
to precess around $\bB_0$.
There are three distinct orientational issues:
\begin{enumerate}
\item The angle $\alpha$ between the grain's
principal axis of largest moment of inertia, $\bahat_1$,
 and the angular momentum $\bJ$ (alignment of the grain body with $\bJ$).
\item The angle $\beta$ between $\bJ$ and $\bB_0$
(alignment of $\bJ$ with $\bB_0$).
\item The angle $\gamma$ between $\bB_0$ and the line-of-sight.
\end{enumerate}
Consider radiation propagating in the $\bzhat$ direction, and suppose
$\bB_0$ to be
in the $\byhat$-$\bzhat$ plane, making an angle
$\gamma$ with the $\bzhat$ axis.
In the electric-dipole limit $a/\lambda\ll 1$, 
the mean absorption cross section and
the polarization cross section
sections for $x$- and $y$-polarized radiation can be written
(see Appendix \ref{app:orientational averages})
\beqa \label{eq:Cx+Cy}
\frac{C_x+C_y}{2}
&\,=\,& \Cran - C_\pol \Phi \left(\sin^2\gamma-\frac{2}{3}\right)
\\ \label{eq:Cx-Cy}
\frac{C_x-C_y}{2} 
&=&C_\pol \Phi \sin^2\gamma
~~~,
\eeqa
where (see Appendix \ref{app:orientational averages})
\beq \label{eq:Phi}
\Phi \equiv 
\frac{9}{4}
\left(\langle\cos^2\alpha\rangle-\frac{1}{3}\right)
\left(\langle\cos^2\beta\rangle-\frac{1}{3}\right)
\eeq
is a generalization of the ``polarization reduction factor''
originally introduced by \citet[][p.\ 328]{Greenberg_1968}
and \citet{Purcell+Spitzer_1971}.
Perfect alignment 
($\langle\cos^2\alpha\rangle=\langle\cos^2\beta\rangle=1$)
has $\Phi=1$;
random orientation ($\langle\cos^2\beta\rangle=1/3$)
results in $\Phi=0$.

If $\bB_0$ is itself not perfectly uniform, 
\citet{Lee+Draine_1985} showed that
$\sin^2\gamma \rightarrow \sin^2\gamma_0 \times
\frac{3}{2}\left(\langle\cos^2\delta\rangle-\frac{1}{3}\right)$
where $\gamma_0$ is now the angle between 
$\bzhat$ and the (dust mass-weighted) mean magnetic field 
$\langle\bB_0\rangle$, and
$\delta$ is the angle between $\langle\bB_0\rangle$ and the local $\bB_0$;
$\langle\cos^2\delta\rangle$ is the dust mass-weighted average of
$\cos^2\delta$ over the sightline.
If we assume that $\alpha$, $\beta$ and $\delta$ vary independently,
then the overall polarization reduction factor becomes
\beq \label{eq:Phi}
\Phi \equiv
\frac{27}{8}
\left(\langle\cos^2\alpha\rangle-\frac{1}{3}\right)
\left(\langle\cos^2\beta\rangle-\frac{1}{3}\right)
\left(\langle\cos^2\delta\rangle-\frac{1}{3}\right)
~~~.
\eeq

Let $N_d$ be the column density of grains, and
$C_x$ and $C_y$ be the average absorption cross section per grain
for radiation
polarized in the $\bxhat$ and $\byhat$ directions.
Let $\tau_x=N_dC_x$ and $\tau_y=N_dC_y$ be the optical depths
for radiation polarized
in the $\bxhat$ and $\byhat$ directions.
Initially unpolarized radiation will be attenuated
and polarized as a
result of linear dichroism (i.e., preferential attenuation of
one linear polarization), with overall
attenuation and fractional polarization
\beqa
I/I_0 &\,=\,& \frac{ e^{-\tau_y}+e^{-\tau_x}}{2}
\\
p &=& \frac{e^{-\tau_y}-e^{-\tau_x}}{e^{-\tau_y}+e^{-\tau_x}}
~~~.
\eeqa
From (\ref{eq:Cx+Cy}) and (\ref{eq:Cx-Cy}) we can find
the absorption cross section per grain volume
$\Cran(\lambda)/V$ from the measured attenuation $I/I_0$ and
polarization $p$ (see Appendix \ref{app:estimating Cran})
where $\rho$ is the mass density of the grain material,
and $\Sigma_{\rm d}$ is the dust mass surface density:
\beqa \label{eq:Cran from tau0}
\frac{\Cran(\lambda)}{V} &\,=\,& 
\frac{\tau_\lambda}{\Sigma_{\rm d}/\rho}
\left[1 +
\frac{p_\lambda}{\tau_\lambda}\left(1-\frac{2}{3\sin^2\gamma}\right)
-\frac{p_\lambda^2}{2\tau_\lambda} - 
\frac{2p_\lambda^3}{3\tau_\lambda}\left(1-\frac{2}{3\sin^2\gamma}\right)
+O\left(\frac{p_\lambda^4}{\tau_\lambda}\right)
\right]
\\
\tau_\lambda &\equiv& \ln\left(I_0/I\right)
~.
\eeqa
Because $p_\lambda/\tau_\lambda$ 
is normally small, common practice is to approximate
$\Cran(\lambda)/V \approx \tau_\lambda/(\Sigma_{\rm d}/\rho)$; 
note, however, that
\citet{Hensley+Zhang+Bock_2019} have demonstrated that the high quality
of the {\it Planck} data permit
the dependence of the total emission on $p_\lambda/\tau_\lambda$ 
(the first-order term in
Eq.\ \ref{eq:Cran from tau0}) to be used to constrain the
full 3D orientation of the magnetic field.

\section{\label{sec:dielectric function}
         Self-Consistent Dielectric Functions Derived from Infrared Absorption}

The relationship between $\Cabs(\lambda)$ and $\Cpol(\lambda)$ 
derived in the previous 
sections can be leveraged on astronomical data in the infrared.
Here we show how the full 
dielectric function $\epsilon(\lambda)$ can be estimated using knowledge of
the infrared opacity.

Suppose that we have an estimate of the dielectric function 
$\epsilon(\lambda)$
of the grain material at short wavelengths $\lambda < \lambda_1$, and have
observational knowledge of the extinction $\tau(\lambda)$ at infrared
wavelengths $\lambda>\lambda_1$, $\Sigma_{\rm d}$, and an estimate for
the grain material density $\rho$.
From these we can estimate the observed absorption
cross section per grain volume $\Cran^{\rm (obs)}/V$ 
for randomly-oriented grains
(see Appendix \ref{app:estimating Cran}).
This applies to the dust material in the ISM, where we have 
constraints on the infrared and far-infrared opacity,
including
the strong silicate absorption features at 9.7$\micron$ and 18$\micron$.
Here we show how one can use the ``observed'' $\Cran^{\rm (obs)}(\lambda)/V$ to
obtain the complex dielectric function
$\epsilon(\lambda)$ at infrared wavelengths.

We assume that at wavelengths $\lambda > \lambda_1$ the grains
have $a\ll\lambda$, so that
we can employ the electric dipole approximation
(\ref{eq:Cran_ed}) 
to relate $\Cran/V$ to the complex dielectric function.
We must, of course, make an assumption about the grain shape, or
distribution of grain shapes.
For spheres, spheroids, ellipsoids, or the CDEs discussed in this paper,
we have analytic expressions relating $\Cran/V$ to the
dielectric function $\epsilon(\lambda)$; the analytic result enables
efficient iterative algorithms to be applied to solve the system of equations.

The dielectric function must satisfy
the Kramers-Kronig relations
\citep{Landau+Lifshitz+Pitaevskii_1993}.
We suppose that we start with a dielectric function
$\epsilon^{0}(\lambda)$ that is reasonably accurate at
$\lambda <\lambda_1$.
We extend the imaginary part of $\epsilon^0$ to long wavelengths
in a smooth way:
\beq
\epsilon_2^0(\lambda)=
\epsilon_2^0(\lambda_1)
\times
\left(\frac{\lambda_1}{\lambda}\right)
~~~,
\eeq
and obtain (by numerical integration) the real part $\epsilon_1^0(\lambda)$
at all wavelengths
using the Kramers-Kronig relation \citep{Landau+Lifshitz+Pitaevskii_1993}:
\beq \label{eq:KK}
\epsilon_1^0(\omega) = 
1 + \frac{2}{\pi}
P
\int_0^\infty \frac{x\epsilon_2^0(x)}{x^2-\omega^2} dx
~~~,
\eeq
where $P$ indicates that the ``principal value'' of the 
singular integral is to be taken.
The actual behavior of $\epsilon_2^0(\lambda>1\micron)$
is unimportant, because we will adjust the total absorption as
required to reproduce $\Cran^{\rm (obs)}/V$ at $\lambda>\lambda_1$.
We accomplish this by adding additional absorption in the form
of $N$ Lorentz oscillators, each with resonant frequency $\omega_{0k}$,
dimensionless damping parameter $\gamma_k$, and dimensionless strength
$S_k$:
\beq \label{eq:epsilon}
\epsilon(\omega) = \epsilon^{(0)}(\omega) + 
\sum_{k=1}^N S_k 
\left[1-\left(\frac{\omega}{\omega_{0k}}\right)^2 
-i\gamma_k\frac{\omega}{\omega_{0k}}\right]^{-1}
~~~.
\eeq
Because $\epsilon^{(0)}(\omega)$ and each of the Lorentz oscillators
separately satisfy the Kramers-Kronig relations,
$\epsilon(\omega)$ given by Eq.\ (\ref{eq:epsilon}) will satisfy the
Kramers-Kronig relations for any $\left\{\omega_{0k},\gamma_k,S_k\right\}$.

We distribute 
the Lorentz oscillators between $\lambda_1$ and $\lambda_N>\lambda_1$ 
according
to some smooth prescription (e.g., uniform in $\log\lambda$).
Then, we set the widths of the Lorentzians by specifying the
dimensionless damping parameters $\gamma_k$: 
\beqa \label{eq:omega_k}
\label{eq:gamma_k}
\gamma_k &\,=\,& C\times \left(\frac{\lambda_j}{\lambda_{j-1}}-1\right)
~~~{\rm with} ~j={\rm min}(2,k)
~~~.
\eeqa
For $\gamma_k\ll 1$, each resonance contributes
Im($\epsilon$) with a FWHM $\approx \gamma_k\omega_{0k}$.
To represent a smooth function, we want
$\gamma_k\omega_{0k}$ to be large compared to 
$\omega_{0,k+1}-\omega_{0k}$, but small enough to be able to reproduce
the expected frequency dependence of ${\rm Im}(\epsilon)$.
This is accomplished by suitable choice for $C$.
For example, \citet{Draine+Hensley_2021a} adopt
$N=3000$, $\omega_{01}/\omega_{0N}=3\cm/1\micron=3\times10^4$,
and $C=10$.

The model cross sections $\Cran^{\rm (model)}(\lambda)$ depend on the
$\{S_k\}$.
To find the self-consistent solution, we 
iteratively adjust the $S_k$ to solve the $N$ simultaneous
equations
\beq
Y_k \equiv \left[\frac{\lambda\Cran^{\rm (model)}}{V}\right]_{\lambda_k}
-
\left[\frac{\lambda\Cran^{\rm (obs)}}{V}\right]_{\lambda_k}
= 0 \hspace*{1.0cm},~~~k=1,...,N
~~~.
\eeq
Thus we have $N$ equations to determine $N$ unknown $S_k$.
Iterative alogrithms, such as the Levenberg-Marquardt method
\citep[see, e.g.,][]{Press+Teukolsky+Vetterling+Flannery_1992}, 
can be used to find the solution $S_k$; it is helpful that 
analytic formulae for the partial derivatives 
$\partial Y_j/\partial S_k$ can be obtained 
from Eq.\ (\ref{eq:epsilon}) and one of
(\ref{eq:Cran_BHCDE}), (\ref{eq:Cran for ERCDE}), or (\ref{eq:Cran for CDE2}).

We remark here that the problem does not always have a solution: if
the ``observed''
$\lambda\Cran^{\rm (obs)}/V$ is too large, there may not be any
dielectric function $\epsilon(\lambda)$ that can reproduce the assumed
$\lambda\Cran^{\rm (obs)}/V$
for the assumed grain shape.  Because of the Kramers-Kronig relations,
all wavelengths matter: strong absorption at one wavelength will imply
a large ${\rm Re}(\epsilon)$ at longer wavelengths, 
limiting the ability of the
grain to absorb at those wavelengths.

We apply this methodology to estimate the effective dielectric
function $\epsilon(\lambda)$
for interstellar dust material in a separate paper 
\citep{Draine+Hensley_2021a}.


\added{
\section{\label{sec:discussion}
         Ellipsoids vs.\ More Complex Shapes}

This paper has concentrated on the optics of grains with spheroidal or
ellipsoidal shapes, including continuous distributions of ellipsoidal
shapes.
In the Rayleigh limit $a\ll\lambda$, the interaction of a grain
with the electromagnetic field
is determined by a single symmetric tensor $\alpha_{jk}$ characterizing the
polarizability of the grain.  For a given dielectric function,
ellipsoidal shapes allow us to explore
plausible values for 
$\alpha_{jk}/V$.

At shorter wavelengths, the response of the grain to an incident
electromagnetic field is more complex, and ellipsoidal shapes provide
only a first approximation to asphericity.
Ellipsoidal shapes may be an adequate approximation for estimation of
cross sections for absorbing or scattering light, for modeling
polarization of starlight at optical wavelengths, or polarized thermal emission
at submm wavelengths.

However, radiative torques are important for grain dynamics, including the
alignment of interstellar grains
\citep{Draine+Weingartner_1996,Draine+Weingartner_1997,Hoang+Lazarian_2008}.
The reflection symmetries (and therefore zero chirality) of ellipsoidal
shapes artificially suppresses radiative torques.  Therefore, studies of 
radiative torques on interstellar grains
must consider non-ellipsoidal grain shapes.  However, the overall
deviations from nonsphericity implied by observations of polarized emission 
at long wavelengths will still serve to constrain the 
more complex shapes used for
studies of starlight torques.

}

\section{\label{sec:summary}
         Summary}

The principal results of this study are as follows:
\begin{enumerate}
\item We discuss the distributions of ellipsoidal shapes that correspond
  to three previously-proposed continuous distributions of ellipsoids (CDEs).
  Twenty randomly-selected shapes from each distribution 
  (Figure \ref{fig:shapes}) serve to illustrate the three distributions.
\item The often-used CDE discussed by \citet{Bohren+Huffman_1983} (here
  referred to as the BHCDE distribution) includes
  what appears to be an unrealistically
  large fraction of extremely elongated or extremely flattened
  shapes.  
\item The CDE2 distribution proposed by
  \citet{Ossenkopf+Henning+Mathis_1992} includes a much smaller fraction
  of extreme shapes, and seems more realistic as a model for
  distributions of grain shapes.
\item For each of the three CDEs considered here, 
  we obtain the distribution functions $g_j(L_j)$
  for the geometric factors $L_1,L_2,L_3$.
\item In the electric dipole limit $a/\lambda\ll 1$, we
  obtain absorption and polarization cross sections for
  partially-aligned
  ellipsoidal grains with the three proposed CDEs.
\item We present a method for obtaining a self-consistent dielectric
  function consistent with an assumed absorption opacity and an
  assumed distribution of shapes.
\end{enumerate}

\acknowledgments
We thank
Eric Stansifer
and
Chris Wright
for helpful discussions.
We thank the referee for helpful comments, and also thank a previous
referee for comments on an earlier version.
This work was supported in part by NSF grants AST-1408723 and AST-1908123,
and carried out in part at the Jet Propulsion Laboratory,
California Institute of Technology, under a contract with
the National Aeronautics and Space Administration.

\appendix

\section{\label{app:CDE}
         Polarization Cross Sections for Grain Populations with
         Continuously-Distributed Ellipticities}

\subsection{General Considerations}

Consider a population of ellipsoids with a distribution of axial ratios.
Every ellipsoidal shape is uniquely specified by its triplet  of
depolarization factors
$(L_1,L_2,L_3)$.  
Because $L_3=1-L_1-L_2$, the ellipsoid is fully-determined by the doublet
$(L_1,L_2)$, which must lie in the triangular region bounded by
$L_1=0$, $L_2=0$, and $L_1+L_2=1$, as shown in Figure \ref{fig:triangle}.

The distribution of shapes can be characterized by the distribution
of $L$ values.
Let $dP=G(L_1,L_2) dL_1dL_2$ be the probability that
$L_1\in (L_1+dL_1)$, $L_2\in (L_2+dL_2)$.
The function $G(L_1,L_2)$ fully determines the shape distribution
(i.e., the distribution of axial ratios).
If $G$ is to apply to the full triangular region in Figure \ref{fig:triangle},
then (because labelling of axes $1,2,3$ is arbitrary), $G$ must
depend symmetrically on $L_1,L_2,L_3$:
\beq \label{eq:Gsym}
G(L_1,L_2)=G(L_2,L_1)=G(L_1,1-L_1-L_2)~~{\rm for~ all~ allowed~} L_1,L_2
~~~.
\eeq

The region of allowed $(L_1,L_2)$ can be divided into 6 triangular
subregions of equal area, shown in Figure \ref{fig:triangle},
corresponding
to the six possible orderings of $L_1,L_2,L_3$:
(1) $L_3\leq L_2\leq L_1$, 
(2) $L_2\leq L_1\leq L_3$, 
(3) $L_1\leq L_3\leq L_2$,
(4) $L_2\leq L_3\leq L_1$, 
(5) $L_3\leq L_1\leq L_2$, and
(6) $L_1\leq L_2\leq L_3$.

For clarity, we fix the order of the
$L$ values:  we choose the ordering 
$0\leq L_3\leq L_2 \leq L_1\leq 1$, corresponding to
region 1 (shaded) in Figure \ref{fig:triangle}.  
Then $L_1$ is for
$\bE$ parallel to the principal axis of largest moment of inertia 
(the ``short axis''),
and $L_3$ is for $\bE$ along the principal axis of smallest
moment of inertia (the ``long axis'').

Within subregion 1, let $g_j(L_j) dL_j$ be the
probability that $L_j\in(L_j,L_j+dL_j)$:
\beq \label{eq:g1g2g3}
\begin{array}{r c l l}
g_1(L_1) &=& 0 & {\rm for~} 0 < L_1 < 1/3 \\
         &=& 6\int_{(1-L_1)/2}^{L_1} G(L_1,L_2) dL_2 & {\rm for~} 1/3 < L_1 < 1/2 \\
         &=& 6\int_{(1-L_1)/2}^{1-L_1} G(L_1,L_2) dL_2 & {\rm for~} 1/2 < L_1 < 1 \\
g_2(L_2) &=& 6\int_{1-2L_2}^{1-L_2} G(L_1,L_2) dL_1  & {\rm for~} 0 < L_2 < 1/3 \\
         &=& 6\int_{L_2}^{1-L_2} G(L_1,L_2) dL_1     & {\rm for~} 1/3 < L_2 < 1/2 \\
         &=& 0                                       & {\rm for~} 1/2 < L_2\\
g_3(L_3) &=& 3\int_0^{1-3L_3} G((1-L_3+z)/2,(1-L_3-z)/2) dz & {\rm for~} 0 < L_3 < 1/3\\
         &=& 0                                       & {\rm for~} 1/3 < L_3 \\
\end{array}
\eeq
where 
we have introduced $z\equiv (L_1-L_2)$ for evaluation of $g_3$.
The factor of six in (\ref{eq:g1g2g3})
appears because we assume the normalization
$\int G dL_1 dL_2=1$ over the full triangular region, hence
$\int G dL_1 dL_2=1/6$ over region 1.
It can be verified that 
\beq
\int_0^1 g_j(L_j) dL_j = 1 ~~~~~{\rm for~} j=1,2,3
~~~.
\eeq
For distributions of ellipsoidal shapes, 
\beq
\langle A_j\rangle \equiv 
\int \frac{(\epsilon -1)}{1+L_j(\epsilon-1)} g_j(L_j) dL_j
~~~.
\eeq

\subsection{BHCDE}

The simplest CDE
is the uniform distribution
\beq \label{eq:original CDE}
G(L_1,L_2)= 2 ~~~{\rm for}~~~ 0\leq L_1+L_2\leq 1
~~~,
\eeq
which obviously satisfies the symmetry condition (\ref{eq:Gsym}).
This example was discussed by \citet{Bohren+Huffman_1983};
we refer to (\ref{eq:original CDE}) as the BHCDE.
For this case we have
\beq
\begin{array}{r c l l}
g_1 &=& 0
&{\rm for}~L_1<\frac{1}{3}
\\
&=& 18(L_1-\frac{1}{3})
&{\rm for}~\frac{1}{3}\leq L_1\leq \frac{1}{2}
\\
&=& 6(1-L_1)
&{\rm for}~\frac{1}{2}\leq L_1 \leq 1
\\
g_2 
&=& 12L_2
&{\rm for}~0 \leq L_2\leq \frac{1}{3}
\\
&=& 12(1-2L_2)
&{\rm for}~\frac{1}{3}\leq L_2\leq \frac{1}{2}
\\
&=& 0
&{\rm for}~\frac{1}{2} \leq L_2
\\
g_3 
&=& 6(1-3L_3)
&{\rm for}~0 \leq L_3\leq \frac{1}{3}
\\
&=& 0
&{\rm for}~\frac{1}{3} \leq L_3
~~.
\end{array}
\eeq
Distributions $g_1$, $g_2$, and $g_3$ are shown in Figure \ref{fig:gj}a.
Then
\beqa
\langle A_1\rangle &\,=\,& 
\frac{6}{x}
\left\{
(1+x)\ln\left(\frac{1+x}{1+x/2}\right)
-3\left(1+\frac{x}{3}\right)\ln\left(\frac{1+x/2}{1+x/3}\right)
\right\}
\\
\langle A_2\rangle &=&
\frac{12}{x}
\left\{
2\left(1+\frac{x}{2}\right)\ln\left(\frac{1+x/2}{1+x/3}\right)
-\ln\left(1+x/3\right)
\right\}
\\
\langle A_3\rangle &=&
\frac{18}{x}\left\{
\left(1+\frac{x}{3}\right)\ln\left(1+x/3\right) - \frac{x}{3}
\right\}
~~.
\\ \label{eq:bhcde <A>}
\frac{\langle A_1+A_2+A_3\rangle}{3}
&=& \frac{2}{x}(1+x)\ln(1+x)-2
~.
\eeqa
Eq.\ (\ref{eq:bhcde <A>}) was previously obtained 
by \citet{Bohren+Huffman_1983}.

\begin{figure}[ht]
\begin{center}
\includegraphics[width=10.0cm,angle=0,clip=true,
                 trim=0.5cm 5.5cm 0.5cm 3.0cm]
{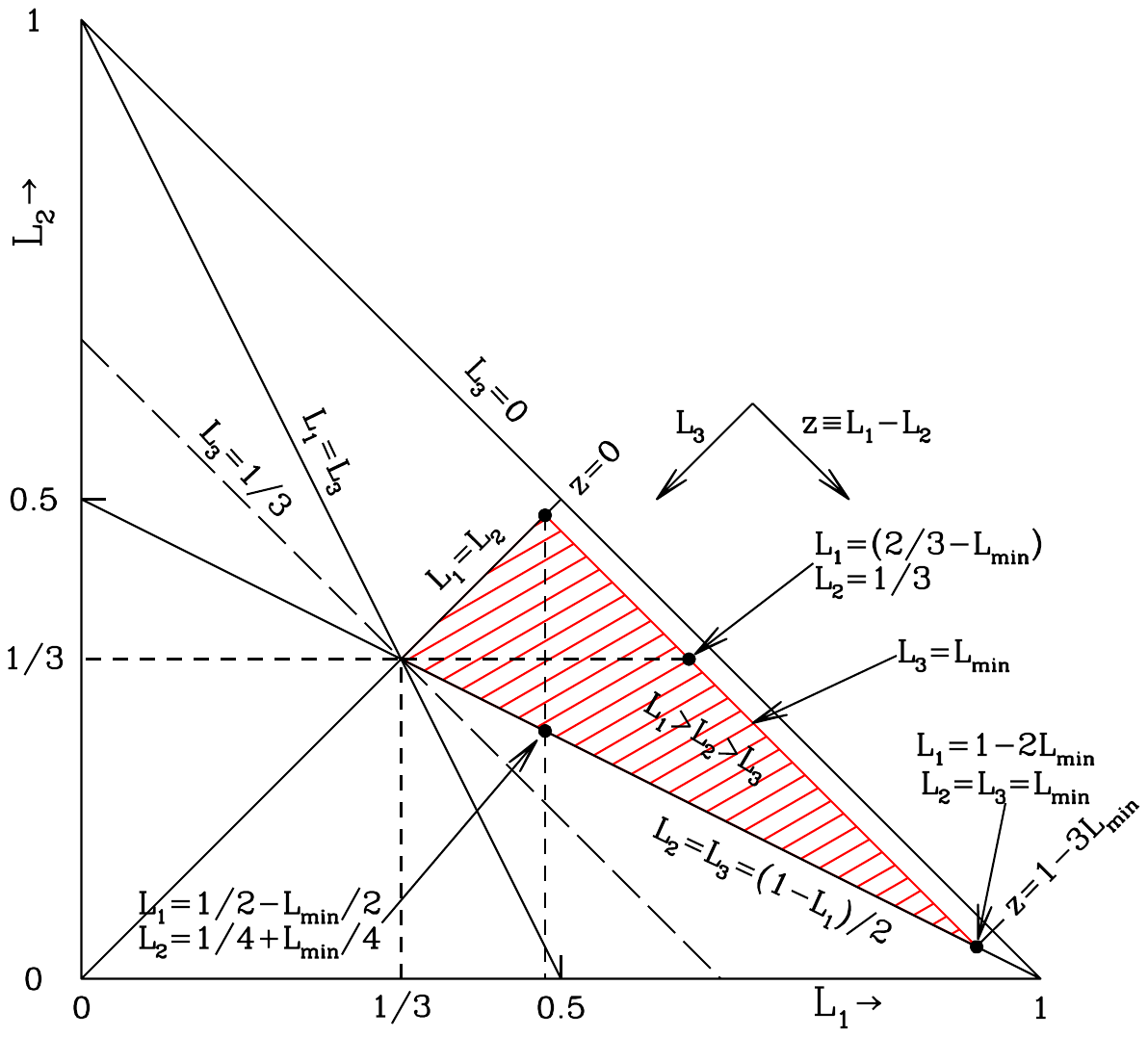}
\caption{\label{fig:ERCDE triangle}
         The shaded area is the ERCDE locus
         with depolarization factors $L_1\geq L_2\geq L_3\geq L_{\min}$
         (see text).}
\end{center}
\end{figure}

\subsection{ERCDE}

The BHCDE includes shapes that
are infinitely elongated ($L_j\rightarrow 0$) and
infinitely flattened ($L_j\rightarrow 1$).
\citet{Zubko+Mennella+Colangeli+Bussoletti_1996} proposed to
exclude the most extreme shapes by imposing the
restriction $L_j\geq \Lmin$, where $0\leq\Lmin\leq1/3$,
giving what Zubko et al.\ referred to as the ``externally restricted
distribution of ellipsoids'' (ERCDE):
\beqa \nonumber
G(L_1,L_2)&\,=\,&\frac{2}{(1-3\Lmin)^2} ~~~{\rm for}\\
\label{eq:ERCDE app}
&& \Lmin \leq L_1 ~,~
\Lmin \leq L_2 ~,~
(L_1+L_2) \leq 1-2\Lmin
~~~.
\eeqa
With $\Lmin=0$ one obtains the original BHCDE; with $\Lmin\rightarrow 1/3$
one obtains spheres.
The domain in the $L_1$--$L_2$ plane is 
shown in Fig.\ \ref{fig:ERCDE triangle}.

\citet{Zubko+Mennella+Colangeli+Bussoletti_1996} obtained 
$\langle A_1+A_2+A_3\rangle$ 
for randomly-oriented grains with the ERCDE
distribution.
Discussion of aligned grains requires the absorption per volume for grains
aligned with the electric fields along their principal axes.
The ERCDE has
\beq
\begin{array}{r c l l}
g_1 &=& 0
&{\rm for}~L_1<\frac{1}{3}
\\
&=& \frac{18}{(1-3\Lmin)^2}\left(L_1-\frac{1}{3}\right)
&{\rm for}~\frac{1}{3}\leq L_1\leq \frac{(1-\Lmin)}{2}
\\
&=& \frac{6}{(1-3\Lmin)^2}\left(1-L_1-2\Lmin\right)
&{\rm for}~\frac{(1-\Lmin)}{2}\leq L_1 \leq 1-2\Lmin
\\
&=& 0
&{\rm for}~1-2\Lmin\leq L_1
\\
g_2 &=& 0
&{\rm for}~L_2<\Lmin
\\
&=& \frac{12}{(1-3\Lmin)^2}(L_2-\Lmin)
&{\rm for}~\Lmin \leq L_2\leq \frac{1}{3}
\\
&=& \frac{12}{(1-3\Lmin)^2}\left(1-\Lmin-2L_2\right)
&{\rm for}~\frac{1}{3}\leq L_2\leq \frac{(1-\Lmin)}{2}
\\
&=& 0
&{\rm for}~\frac{(1-\Lmin)}{2} \leq L_2
\\
g_3 &=& 0
&{\rm for}~L_3\leq\Lmin
\\
&=& \frac{6}{(1-3\Lmin)^2}(1-3L_3)
&{\rm for}~\Lmin \leq L_3\leq \frac{1}{3}
\\
&=& 0
&{\rm for}~\frac{1}{3} \leq L_3
~~.
\end{array}
\eeq
Distributions $g_1$, $g_2$, and $g_3$ are shown in Figure \ref{fig:gj}b for
$\Lmin=0.05$.
It is convenient to define
\beqa
B &\,\equiv\,& 1/2-\Lmin/2
\\
D &\equiv& 1-2\Lmin
\\
x &\equiv& \epsilon-1
~~~.
\eeqa
We obtain
\beqa
\langle A_1\rangle &\,=\,& \frac{6}{(1-3\Lmin)^2}
\left\{
\left(\frac{1}{x} + D\right)
      \ln\left[\frac{1+ xD}{1+xB}\right]
-3\left(\frac{1}{x} + \frac{1}{3}\right)
      \ln\left[\frac{1+ xB}{1+x/3}\right]
\right\}
\\
\langle A_2\rangle &=& \frac{12}{(1-3\Lmin)^2}
\left\{
2\left(\frac{1}{x} + B\right)
\ln\left[\frac{1+xB}{1+x/3}\right]
-\left(\frac{1}{x}+\Lmin\right)
\ln\left[\frac{1+x/3}{1+x\Lmin}\right]
\right\}~~~~
\\
\langle A_3\rangle &=& \frac{18}{(1-3\Lmin)^2}
\left\{
\left( \frac{1}{x}+\frac{1}{3} \right)
\ln\left[\frac{1 + x/3}{1+x\Lmin}\right]
-\frac{(1-3\Lmin)}{3} 
\right\}
\\ \label{eq:ercde <A>}
\frac{\langle A_1+A_2+A_3\rangle}{3}
&=&
\frac{2}{(1-3\Lmin)^2}
\left\{\left(\frac{1}{x}+
D\right)\ln\left[\frac{1+xD}{1+x\Lmin}\right]
-(1-3\Lmin)\right\}
~.
\eeqa
Eq.\ (\ref{eq:ercde <A>}) was previously obtained by
\citet{Zubko+Mennella+Colangeli+Bussoletti_1996}.

\subsection{CDE2}

\citet{Ossenkopf+Henning+Mathis_1992} proposed
the distribution
\beq
G(L_1,L_2)=120 L_1 L_2 L_3
~~~.
\eeq
This satifies the symmetry requirement (\ref{eq:Gsym}), and
has the desirable property that $G\rightarrow0$ for
$L_j\rightarrow 0$.
We find
\beq \label{eq:gj for new CDE}
\begin{array}{r c l l}
g_1(L_1) &=& 0
&{\rm for~} L_1< \frac{1}{3}
\\
&=& 60L_1\left(-1 +3L_1 +3L_1^2 -9 L_1^3\right) 
&{\rm for~} \frac{1}{3} \leq L_1\leq \frac{1}{2}
\\
         &=& 60 L_1 (1-L_1)^3
& {\rm for~} \frac{1}{2} \leq L_1\leq 1
\\
g_2(L_2) &=& 120 L_2^3 (3-5L_2)
&{\rm for~} 0 \leq L_2\leq \frac{1}{3}
\\
         &=& 120(L_2-3L_2^2+4L_2^4)
&{\rm for~} \frac{1}{3} \leq L_2\leq \frac{1}{2}
\\
&=& 0
&{\rm for~} \frac{1}{2} < L_2
\\
g_3(L_3) &=& 60 L_3(1 - 3L_3 -3L_3^2 + 9L_3^3)
&{\rm for~} 0 \leq L_3\leq \frac{1}{3}
\\
&=& 0
&{\rm for~} \frac{1}{3} < L_3
~~~.
\end{array}
\eeq
These distributions are shown in Figure \ref{fig:gj}c.
To have a sense of how nonspherical a typical ellipsoid from
this distribution might be, we consider the mean
depolarization factors 
$\langle L_j\rangle\equiv\int L_j g_j(L_j)dL_j$.
For $g_j$ given by eq.\ (\ref{eq:gj for new CDE}) we
find
$\langle L_1\rangle=0.5355$, $\langle L_2\rangle= 0.3040$,
and $\langle L_3\rangle=0.1605$.
These mean values correspond to an ellipsoid with axial ratios
$a_1:a_2:a_3::1:1.664:2.716$.

For $g_j$ given by eq.\ (\ref{eq:gj for new CDE}) we obtain
\beqa \nonumber
\langle A_1 \rangle &\,=\,& \frac{60}{x^4}\bigg[
(9+3x-3x^2-x^3)\ln\left(1+\frac{x}{3}\right)
+(-8+6x^2+2x^3)\ln\left(1+\frac{x}{2}\right)
\\
&&~~~~~~~
+(-1-3x-3x^2-x^3)\ln\left(1+x\right)
+2x+x^2+\frac{2}{9}x^3+\frac{7}{216}x^4
\bigg]
\\ \nonumber
\langle A_2 \rangle &=& \frac{60}{x^4}\bigg[
(-18-6x+6x^2+2x^3)\ln\left(1+\frac{x}{3}\right)
+(8-6x^2-2x^3)\ln\left(1+\frac{x}{2}\right)
\\
&&~~~~~~~
+2x+2x^2+\frac{5}{9}x^3+\frac{13}{216}x^4\bigg]
\\
\langle A_3 \rangle &=& \frac{60}{x^4}\bigg[
(9+3x-3x^2-x^3)\ln\left(1+\frac{x}{3}\right)
-3x-\frac{1}{2}x^2+\frac{19}{18}x^3 + \frac{17}{108}x^4\bigg]
\\ \label{eq:cde2 <A>}
\frac{\langle A_1+A_2+A_3\rangle}{3}
&=&
\frac{20}{x^4}\left[-(1+x)^3\ln(1+x)+x+\frac{5}{2}x^2
+\frac{11}{6}x^3+\frac{1}{4}x^4\right]
~,
\eeqa
where $x\equiv \epsilon-1$.
Eq.\ (\ref{eq:cde2 <A>}) was previously obtained by
\citet{Fabian+Henning+Jager+etal_2001}.

\section{\label{app:orientational averages}
         Orientation-Averaged Cross Sections for Partially-Aligned
Grains}
Consider radiation propagating along the $\bzhat$ axis.
Let the local magnetic field be in the $\byhat-\bzhat$ plane,
with $\gamma=$ the angle between $\bB$ and the line-of-sight:
$\bBhat=\byhat\sin\gamma+\bzhat\cos\gamma$.
Let $\bJhat$ be a unit vector in the direction of the grain's angular
momentum, and $\beta$ the angle between $\bJhat$ and $\bBhat$.
If $\beta> 0$, 
the grain's magnetic moment will cause $\bJhat$ to precess
around $\bBhat$, and we may write
\beqa
\bJhat &\,=\,& \bBhat\cos\beta + \bxhat\sin\beta\cos\phi_1
+ (\bxhat\times\bBhat)\sin\beta\sin\phi_1
\\
&=& \bxhat\sin\beta\cos\phi_1 + \byhat(\cos\beta\sin\gamma-\sin\beta\cos\gamma\sin\phi_1) + \bzhat(\cos\beta\cos\gamma+\sin\beta\sin\gamma\sin\phi_1)
~,~~
\eeqa
with $\phi_1$ varying from $0$ to $2\pi$ over one precession period.
Observations of starlight polarization 
indicate that there is systematic alignment of
$\bJ$ with $\bB$, i.e., $\langle\cos^2\beta\rangle>1/3$,
with the alignment presumed to result from some combination of
paramagnetic dissipation \citep{Davis+Greenstein_1951},
superparamagnetic dissipation \citep{Jones+Spitzer_1967},
ferromagnetic dissipation \citep{Draine+Hensley_2013}
or starlight torques
\citep{Draine+Weingartner_1997,Weingartner+Draine_2003,
       Hoang+Lazarian_2009a,Hoang+Lazarian_2009b}.

On short time scales the grain spins and nutates with fixed $\bJhat$
according to the dynamics of rigid bodies
\citep[see, e.g.,][]{Weingartner+Draine_2003}.
Let $\bahat_1$ be the principal axis of
largest moment of inertia, and
let $\alpha$ be the angle between $\bJhat$ and $\bahat_1$.
At constant $J$ and kinetic energy $E_{\rm rot}$ the grain will
tumble: $\bahat$ will nutate around
$\bJhat$.  If the grain is triaxial, the angle $\alpha$ does
not remain constant during the nutation, but will have
some time-averaged value of $\langle \cos^2\alpha\rangle$.

For fixed $J$, the kinetic energy of the grain 
is minimized
if $\alpha=0$ ($\cos^2\alpha=1$).
If the direction of $\bahat$ is uncorrelated with $\bJhat$, then
$\langle\cos^2\alpha\rangle=1/3$.
Thus we expect dissipation in the grain to result in
$\langle\cos^2\alpha\rangle>1/3$.
Suprathermally rotating grains, with rotational kinetic energy
$E_{\rm rot} \gg kT_{\rm grain}$, are expected to have
$\cos^2\alpha\approx 1$ as the result of dissipation associated
with viscoelasticity \citep{Purcell_1979} or the even greater
dissipation associated with the Barnett effect
\citep{Lazarian+Roberge_1997} and nuclear spin
relaxation \citep{Lazarian+Draine_1999b}.

After averaging over precession and nutation,
\beqa
\langle(\bahat_1\cdot\bxhat)^2\rangle &\,=\,&
\frac{1}{3} - 
\frac{3}{4}
\left(\langle\cos^2\alpha\rangle-\frac{1}{3}\right)
\left(\cos^2\beta-\frac{1}{3}\right)
\\
\langle(\bahat_1\cdot\byhat)^2\rangle &=&
\frac{1}{3}
+\frac{9}{4}
\left(\langle\cos^2\alpha\rangle-\frac{1}{3}\right)
\left(\cos^2\beta-\frac{1}{3}\right)
\left(\sin^2\gamma-\frac{1}{3}\right)
\\
\langle(\bahat_2\cdot\bxhat)^2\rangle =
\langle(\bahat_3\cdot\bxhat)^2\rangle &=& 
\frac{1}{3} + 
\frac{3}{8}\left(\langle\cos^2\alpha\rangle-\frac{1}{3}\right)
\left(\langle\cos^2\beta\rangle-\frac{1}{3}\right)
\\
\langle(\bahat_2\cdot\byhat)^2\rangle =
\langle(\bahat_3\cdot\byhat)^2\rangle &=&
\frac{1}{3} - 
\frac{9}{8}
\left(\langle\cos^2\alpha\rangle-\frac{1}{3}\right)
\left(\cos^2\beta-\frac{1}{3}\right)
\left(\sin^2\gamma-\frac{1}{3}\right)
~~~.
\eeqa
The cross sections for radiation polarized in the $\bxhat$ and
$\byhat$ directions are
\beqa
C_x &\,=\,& \Cran + \frac{2}{3} \Cpol \Phi
\\
C_y &=& \Cran - 2\Cpol\Phi\left(\sin^2\gamma -\frac{1}{3}\right)
\\ \label{eq:b1}
\frac{C_x+C_y}{2} &=&
\Cran -\Cpol\Phi\left(\sin^2\gamma-\frac{2}{3}\right)
\\ \label{eq:b2}
\frac{C_x-C_y}{2} &=&
\Cpol\Phi\sin^2\gamma
\\ \label{eq:b3}
\Cran&\equiv& \frac{1}{3}\left[
\Cabs(\bE\parallel\bahat_1)+
\Cabs(\bE\parallel\bahat_2)+
\Cabs(\bE\parallel\bahat_3)
\right]
\\ \label{eq:b4}
\Cpol&\equiv& \frac{1}{4}
\left[
\Cabs(\bE\parallel\bahat_2)+
\Cabs(\bE\parallel\bahat_3)-
2\Cabs(\bE\parallel\bahat_1)
\right]
\\
\Phi &\equiv& \frac{9}{4}
\left(\langle\cos^2\alpha\rangle-\frac{1}{3}\right)
\left(\cos^2\beta-\frac{1}{3}\right)
~~~.
\eeqa

\section{\label{app:estimating Cran}
         Estimating $\Cran$ from Observations}

Suppose that the attenuation
$I/I_0$ is known,
where the intensity $I(\lambda)$ is summed over both polarization
modes, and the unattenuated radiation $I_0$ is 
unpolarized.
The fractional
polarization $p(\lambda)$ is also measured.
Let $\bzhat$ be the direction of propagation, 
and $\byhat$ be the polarization direction.
If $N$ is the total column density of grains, we
seek to determine the cross section $\Cran(\lambda)$ 
for randomly-oriented grains.
Define
\beqa
\bar{\tau}&\,\equiv\,& \frac{\tau_x+\tau_y}{2}
\\
\tau_p&\equiv& \frac{\tau_x-\tau_y}{2}
~~.
\eeqa
Then
\beqa
\frac{I}{I_0} &\,=\,& \frac{e^{-\tau_x}+e^{-\tau_y}}{2}
\\
&=&e^{-\bar{\tau}}\left[1-\frac{1}{2}\tau_p^2 + O(\tau_p^4)\right]
~~.
\\
p &=& \frac{e^{-\tau_y}-e^{-\tau_x}}{2I/I_0}
\\
&=& \tau_p\frac{\left[1+\frac{1}{6}\tau_p^2 + O(\tau_p^4)\right]}
{\left[1-\frac{1}{2}\tau_p^2 + O(\tau_p^4)\right]}
= \tau_p\left[1+\frac{2}{3}\tau_p^2 + O(\tau_p^4)\right]
~~,
\\ \label{eq:C45}
\tau_p&\approx& p - \frac{2}{3}p^3 + O(p^5)
~~,
\\
\bar{\tau} &=& \ln(I_0/I) + 
\ln\left[1-\frac{1}{2}\tau_p^2 + O(\tau_p^4)\right]
\\
&\approx& \ln(I_0/I) -\frac{1}{2}\tau_p^2 + O(\tau_p^4)
\\ \label{eq:C48}
&\approx& \ln(I_0/I) -\frac{1}{2}p^2 + O(p^4)
~~~.
\eeqa
From (\ref{eq:b1}--\ref{eq:b4}) we have
\beqa
\bar{\tau} &\,=\,& 
N\left[\Cran - \Cpol\Phi\left(\sin^2\gamma-\frac{2}{3}\right)\right]
\\
&=& N\Cran - \tau_p\left(1-\frac{2}{3\sin^2\gamma}\right)
~~.
\eeqa
Using (\ref{eq:C45}) and (\ref{eq:C48}) we obtain
\beqa
\Cran &\,=\,& \frac{1}{N}
\left[\ln\left(\frac{I_0}{I}\right) + p\left(1-\frac{2}{3\sin^2\gamma}\right)
-\frac{1}{2}p^2 - \frac{2}{3}p^3\left(1-\frac{2}{3\sin^2\gamma}\right)
+ O(p^4)\right]
~~.~~~
\eeqa
If the polarization fraction $p\ll 1$, we may approximate
$\Cran\approx (1/N)\ln(I_0/I)$.  For finite $p\ltsim0.2$, we can correct
for the alignment if $p$ is measured and $\sin^2\gamma$
can be estimated.

\section{\label{app:one-to-one mapping}
         Proof of Uniqueness}

For an ellipsoid with semi-major axes $a_1 \leq a_2 \leq a_3$, the
corresponding shape factors $L_1 \geq L_2 \geq L_3$ are given by
Eq.\ (\ref{eq:L_j from a_j},\ref{eq:y_j}).
While we do not offer a proof that there is an $(a_2/a_1,a_3/a_1)$
corresponding to every possible $(L_1,L_2,L_3)$, we have implemented
a numerical procedure that always returns a solution.
In this note, we demonstrate that this solution is unique. 

Suppose that $(a_2/a_1,a_3/a_1)$ corresponds to the desired
$(L_1, L_2, L_3)$.
Without loss of generality, let $a_1 =
1$.  We may then rewrite
\begin{equation}
L_j = \frac{a_2 a_3}{2} \int_0^\infty \frac{dx}{\left(a_j^2 +
    x\right)\left[\left(1+x\right)\left(a_2^2 + x\right)\left(a_3^2 +
      x\right)\right]^{1/2}}
~~~.
\end{equation}
Computing the derivatives
\begin{align}
\frac{\partial L_1}{\partial a_2} &= \frac{1}{2} \int_0^\infty
                                    \frac{a_3 x dx}{\left(1 +
                                    x\right)^{3/2}\left(a_2^2 +
                                    x\right)^{3/2}\left(a_3^2 +
                                    x\right)^{1/2}} \\
\frac{\partial L_1}{\partial a_3} &= \frac{1}{2} \int_0^\infty
                                    \frac{a_2 x dx}{\left(1 +
                                    x\right)^{3/2}\left(a_2^2 +
                                    x\right)^{1/2}\left(a_3^2 +
                                    x\right)^{3/2}} \\
\frac{\partial L_2}{\partial a_3} &= \frac{1}{2} \int_0^\infty
                                    \frac{a_2 x dx}{\left(1 +
                                    x\right)^{1/2}\left(a_2^2 +
                                    x\right)^{3/2}\left(a_3^2 +
                                    x\right)^{3/2}} \\
\frac{\partial L_3}{\partial a_2} &= \frac{1}{2} \int_0^\infty
                                    \frac{a_3 x dx}{\left(1 +
                                    x\right)^{1/2}\left(a_2^2 +
                                    x\right)^{3/2}\left(a_3^2 +
                                    x\right)^{3/2}}
~~~,
\end{align}
we see that the integrands are positive definite for all $a_2$, $a_3$,
and $x$. Therefore,
\beq
\frac{\partial L_1}{\partial a_2} > 0 ~, ~~~
\frac{\partial L_1}{\partial a_3} > 0 ~, ~~~
\frac{\partial L_2}{\partial a_3} > 0 ~, ~~~
\frac{\partial L_3}{\partial a_2} > 0 ~.
\eeq




Because the $L_j$ sum to one, it must be true that
\begin{align}
\frac{\partial L_1}{\partial a_2} + \frac{\partial L_2}{\partial a_2}
  + \frac{\partial L_3}{\partial a_2}&= 0 \\
\frac{\partial L_1}{\partial a_3} + \frac{\partial L_2}{\partial a_3}
  + \frac{\partial L_3}{\partial a_3}&= 0
~~~,
\end{align}
and so
\beq
 \frac{\partial L_2}{\partial a_2} < 0 ~~~,~~~
\frac{\partial L_3}{\partial a_3} < 0
~~~.
\eeq

Assume that there are two sets of axial ratios $(a_2, a_3)$ and
$(a_2', a_3')$ which yield the same $(L_1, L_2, L_3)$. We will proceed
by starting from $(a_2, a_3)$ and adjusting the axial ratios one at a
time to the values $(a_2', a_3')$. We will show that it is impossible
to make a nonzero adjustment and return back to the original $(L_1,
L_2, L_3)$. Note that since $1 \leq a_2 \leq a_3$ by construction,
$L_1 \geq L_2 \geq L_3$ and thus permutations of the $L_j$ are
excluded.

If $a_2 > a_2'$, we can first decrease $a_2$ until it is equal to $a_2'$. From
the realtions above, doing so decreases $L_1$, increases $L_2$, and
decreases $L_3$. To return the $L_j$ to their original values,
adjusting $a_3$ must increase $L_1$, decrease $L_2$, and increase
$L_3$. However, decreasing $a_3$ decreases $L_1$ while increasing
$a_3$ increases $L_2$, and so the desired adjustment is not
possible. An analogous argument holds for $a_2 < a_2'$.

Therefore, $(a_2, a_3)$ is the {\it unique} set of axial ratios
corresponding to $(L_1, L_2, L_3)$.

\bibliography{/u/draine/work/libe/btdrefs}

\end{document}